\documentclass[longauth]{aa}

\usepackage{graphicx}
\usepackage{natbib}
\usepackage{scalerel}

\usepackage[table]{xcolor}

\usepackage{txfonts}
\usepackage[pdfencoding=auto,psdextra]{hyperref}
\hypersetup{
    colorlinks=true,
    linkcolor=blue,
    filecolor=magenta,      
    urlcolor=blue,
    citecolor=blue
}
\makeatletter
\renewcommand*\aa@pageof{, page \thepage{} of \pageref*{LastPage}}
\makeatother

\usepackage[utf8]{inputenc}

\usepackage[switch, modulo]{lineno}
\linenumbers

\usepackage{euclid}

\begin{document}
\nolinenumbers   
%
%
\title{Euclid: Early Release Observations -- Unveiling the morphology of two Milky Way globular clusters out to their periphery\thanks{This paper is published on behalf of the Euclid Consortium.}}    

\author{D.~Massari\orcid{0000-0001-8892-4301}\thanks{\email{davide.massari@inaf.it}}\inst{\ref{aff1}}
\and E.~Dalessandro\orcid{0000-0003-4237-4601}\inst{\ref{aff1}}
\and D.~Erkal\orcid{0000-0002-8448-5505}\inst{\ref{aff2}}
\and E.~Balbinot\orcid{0000-0002-1322-3153}\inst{\ref{aff3},\ref{aff4}}
\and J.~Bovy\orcid{0000-0001-6855-442X}\inst{\ref{aff5}}
\and I.~McDonald\orcid{0000-0003-0356-0655}\inst{\ref{aff6}}
\and A.~M.~N.~Ferguson\inst{\ref{aff7}}
\and S.~S.~Larsen\orcid{0000-0003-0069-1203}\inst{\ref{aff8}}
\and A.~Lan\c{c}on\orcid{0000-0002-7214-8296}\inst{\ref{aff9}}
\and F.~Annibali\inst{\ref{aff1}}
\and B.~Goldman\orcid{0000-0002-2729-7276}\inst{\ref{aff10},\ref{aff9}}
\and P.~B.~Kuzma\orcid{0000-0003-1980-8838}\inst{\ref{aff7},\ref{aff11}}
\and K.~Voggel\orcid{0000-0001-6215-0950}\inst{\ref{aff12}}
\and T.~Saifollahi\orcid{0000-0002-9554-7660}\inst{\ref{aff4},\ref{aff9}}
\and J.-C.~Cuillandre\orcid{0000-0002-3263-8645}\inst{\ref{aff13}}
\and M.~Schirmer\orcid{0000-0003-2568-9994}\inst{\ref{aff14}}
\and M.~Kluge\orcid{0000-0002-9618-2552}\inst{\ref{aff15}}
\and B.~Altieri\orcid{0000-0003-3936-0284}\inst{\ref{aff16}}
\and A.~Amara\inst{\ref{aff2}}
\and S.~Andreon\orcid{0000-0002-2041-8784}\inst{\ref{aff17}}
\and N.~Auricchio\orcid{0000-0003-4444-8651}\inst{\ref{aff1}}
\and M.~Baldi\orcid{0000-0003-4145-1943}\inst{\ref{aff18},\ref{aff1},\ref{aff19}}
\and A.~Balestra\orcid{0000-0002-6967-261X}\inst{\ref{aff20}}
\and S.~Bardelli\orcid{0000-0002-8900-0298}\inst{\ref{aff1}}
\and A.~Basset\inst{\ref{aff21}}
\and R.~Bender\orcid{0000-0001-7179-0626}\inst{\ref{aff15},\ref{aff22}}
\and D.~Bonino\inst{\ref{aff23}}
\and E.~Branchini\orcid{0000-0002-0808-6908}\inst{\ref{aff24},\ref{aff25},\ref{aff17}}
\and M.~Brescia\orcid{0000-0001-9506-5680}\inst{\ref{aff26},\ref{aff27},\ref{aff28}}
\and J.~Brinchmann\orcid{0000-0003-4359-8797}\inst{\ref{aff29}}
\and S.~Camera\orcid{0000-0003-3399-3574}\inst{\ref{aff30},\ref{aff31},\ref{aff23}}
\and G.~P.~Candini\orcid{0000-0001-9481-8206}\inst{\ref{aff32}}
\and V.~Capobianco\orcid{0000-0002-3309-7692}\inst{\ref{aff23}}
\and C.~Carbone\orcid{0000-0003-0125-3563}\inst{\ref{aff33}}
\and R.~G.~Carlberg\inst{\ref{aff5}}
\and J.~Carretero\orcid{0000-0002-3130-0204}\inst{\ref{aff34},\ref{aff35}}
\and S.~Casas\orcid{0000-0002-4751-5138}\inst{\ref{aff36}}
\and M.~Castellano\orcid{0000-0001-9875-8263}\inst{\ref{aff37}}
\and S.~Cavuoti\orcid{0000-0002-3787-4196}\inst{\ref{aff27},\ref{aff28}}
\and A.~Cimatti\inst{\ref{aff38}}
\and G.~Congedo\orcid{0000-0003-2508-0046}\inst{\ref{aff7}}
\and C.~J.~Conselice\orcid{0000-0003-1949-7638}\inst{\ref{aff6}}
\and L.~Conversi\orcid{0000-0002-6710-8476}\inst{\ref{aff39},\ref{aff16}}
\and Y.~Copin\orcid{0000-0002-5317-7518}\inst{\ref{aff40}}
\and L.~Corcione\orcid{0000-0002-6497-5881}\inst{\ref{aff23}}
\and F.~Courbin\orcid{0000-0003-0758-6510}\inst{\ref{aff41}}
\and H.~M.~Courtois\orcid{0000-0003-0509-1776}\inst{\ref{aff42}}
\and H.~Degaudenzi\orcid{0000-0002-5887-6799}\inst{\ref{aff43}}
\and J.~Dinis\inst{\ref{aff44},\ref{aff45}}
\and F.~Dubath\orcid{0000-0002-6533-2810}\inst{\ref{aff43}}
\and X.~Dupac\inst{\ref{aff16}}
\and S.~Dusini\orcid{0000-0002-1128-0664}\inst{\ref{aff46}}
\and M.~Fabricius\inst{\ref{aff15},\ref{aff22}}
\and M.~Farina\orcid{0000-0002-3089-7846}\inst{\ref{aff47}}
\and S.~Farrens\orcid{0000-0002-9594-9387}\inst{\ref{aff48}}
\and S.~Ferriol\inst{\ref{aff40}}
\and M.~Frailis\orcid{0000-0002-7400-2135}\inst{\ref{aff49}}
\and E.~Franceschi\orcid{0000-0002-0585-6591}\inst{\ref{aff1}}
\and B.~Garilli\orcid{0000-0001-7455-8750}\inst{\ref{aff33}}
\and B.~Gillis\orcid{0000-0002-4478-1270}\inst{\ref{aff7}}
\and C.~Giocoli\orcid{0000-0002-9590-7961}\inst{\ref{aff1},\ref{aff50}}
\and A.~Grazian\orcid{0000-0002-5688-0663}\inst{\ref{aff20}}
\and L.~Guzzo\orcid{0000-0001-8264-5192}\inst{\ref{aff51},\ref{aff17},\ref{aff52}}
\and J.~Hoar\inst{\ref{aff16}}
\and H.~Hoekstra\orcid{0000-0002-0641-3231}\inst{\ref{aff3}}
\and M.~S.~Holliman\inst{\ref{aff53}}
\and W.~Holmes\inst{\ref{aff54}}
\and I.~Hook\orcid{0000-0002-2960-978X}\inst{\ref{aff55}}
\and F.~Hormuth\inst{\ref{aff56}}
\and A.~Hornstrup\orcid{0000-0002-3363-0936}\inst{\ref{aff57},\ref{aff58}}
\and P.~Hudelot\inst{\ref{aff59}}
\and K.~Jahnke\orcid{0000-0003-3804-2137}\inst{\ref{aff14}}
\and E.~Keih\"anen\orcid{0000-0003-1804-7715}\inst{\ref{aff60}}
\and S.~Kermiche\orcid{0000-0002-0302-5735}\inst{\ref{aff61}}
\and A.~Kiessling\orcid{0000-0002-2590-1273}\inst{\ref{aff54}}
\and T.~Kitching\orcid{0000-0002-4061-4598}\inst{\ref{aff32}}
\and R.~Kohley\inst{\ref{aff16}}
\and B.~Kubik\orcid{0009-0006-5823-4880}\inst{\ref{aff40}}
\and M.~K\"ummel\orcid{0000-0003-2791-2117}\inst{\ref{aff22}}
\and M.~Kunz\orcid{0000-0002-3052-7394}\inst{\ref{aff62}}
\and H.~Kurki-Suonio\orcid{0000-0002-4618-3063}\inst{\ref{aff63},\ref{aff64}}
\and S.~Ligori\orcid{0000-0003-4172-4606}\inst{\ref{aff23}}
\and P.~B.~Lilje\orcid{0000-0003-4324-7794}\inst{\ref{aff65}}
\and V.~Lindholm\orcid{0000-0003-2317-5471}\inst{\ref{aff63},\ref{aff64}}
\and I.~Lloro\inst{\ref{aff66}}
\and D.~Maino\inst{\ref{aff51},\ref{aff33},\ref{aff52}}
\and E.~Maiorano\orcid{0000-0003-2593-4355}\inst{\ref{aff1}}
\and O.~Mansutti\orcid{0000-0001-5758-4658}\inst{\ref{aff49}}
\and O.~Marggraf\orcid{0000-0001-7242-3852}\inst{\ref{aff67}}
\and K.~Markovic\orcid{0000-0001-6764-073X}\inst{\ref{aff54}}
\and N.~Martinet\orcid{0000-0003-2786-7790}\inst{\ref{aff68}}
\and F.~Marulli\orcid{0000-0002-8850-0303}\inst{\ref{aff69},\ref{aff1},\ref{aff19}}
\and R.~Massey\orcid{0000-0002-6085-3780}\inst{\ref{aff70}}
\and S.~Maurogordato\inst{\ref{aff71}}
\and E.~Medinaceli\orcid{0000-0002-4040-7783}\inst{\ref{aff1}}
\and S.~Mei\orcid{0000-0002-2849-559X}\inst{\ref{aff72}}
\and Y.~Mellier\inst{\ref{aff73},\ref{aff59}}
\and M.~Meneghetti\orcid{0000-0003-1225-7084}\inst{\ref{aff1},\ref{aff19}}
\and G.~Meylan\inst{\ref{aff41}}
\and M.~Moresco\orcid{0000-0002-7616-7136}\inst{\ref{aff69},\ref{aff1}}
\and L.~Moscardini\orcid{0000-0002-3473-6716}\inst{\ref{aff69},\ref{aff1},\ref{aff19}}
\and E.~Munari\orcid{0000-0002-1751-5946}\inst{\ref{aff49}}
\and R.~Nakajima\inst{\ref{aff67}}
\and R.~C.~Nichol\inst{\ref{aff2}}
\and S.-M.~Niemi\inst{\ref{aff74}}
\and C.~Padilla\orcid{0000-0001-7951-0166}\inst{\ref{aff34}}
\and S.~Paltani\orcid{0000-0002-8108-9179}\inst{\ref{aff43}}
\and F.~Pasian\orcid{0000-0002-4869-3227}\inst{\ref{aff49}}
\and K.~Pedersen\inst{\ref{aff75}}
\and W.~J.~Percival\orcid{0000-0002-0644-5727}\inst{\ref{aff76},\ref{aff77},\ref{aff78}}
\and V.~Pettorino\inst{\ref{aff74}}
\and S.~Pires\orcid{0000-0002-0249-2104}\inst{\ref{aff48}}
\and G.~Polenta\orcid{0000-0003-4067-9196}\inst{\ref{aff79}}
\and M.~Poncet\inst{\ref{aff21}}
\and L.~A.~Popa\inst{\ref{aff80}}
\and L.~Pozzetti\orcid{0000-0001-7085-0412}\inst{\ref{aff1}}
\and G.~D.~Racca\inst{\ref{aff74}}
\and F.~Raison\orcid{0000-0002-7819-6918}\inst{\ref{aff15}}
\and R.~Rebolo\inst{\ref{aff81},\ref{aff82}}
\and A.~Renzi\orcid{0000-0001-9856-1970}\inst{\ref{aff83},\ref{aff46}}
\and J.~Rhodes\inst{\ref{aff54}}
\and G.~Riccio\inst{\ref{aff27}}
\and Hans-Walter~Rix\orcid{0000-0003-4996-9069}\inst{\ref{aff14}}
\and E.~Romelli\orcid{0000-0003-3069-9222}\inst{\ref{aff49}}
\and M.~Roncarelli\orcid{0000-0001-9587-7822}\inst{\ref{aff1}}
\and E.~Rossetti\inst{\ref{aff18}}
\and R.~Saglia\orcid{0000-0003-0378-7032}\inst{\ref{aff22},\ref{aff15}}
\and D.~Sapone\orcid{0000-0001-7089-4503}\inst{\ref{aff84}}
\and B.~Sartoris\inst{\ref{aff22},\ref{aff49}}
\and P.~Schneider\orcid{0000-0001-8561-2679}\inst{\ref{aff67}}
\and T.~Schrabback\orcid{0000-0002-6987-7834}\inst{\ref{aff85}}
\and A.~Secroun\orcid{0000-0003-0505-3710}\inst{\ref{aff61}}
\and G.~Seidel\orcid{0000-0003-2907-353X}\inst{\ref{aff14}}
\and M.~Seiffert\orcid{0000-0002-7536-9393}\inst{\ref{aff54}}
\and S.~Serrano\orcid{0000-0002-0211-2861}\inst{\ref{aff86},\ref{aff87},\ref{aff88}}
\and C.~Sirignano\orcid{0000-0002-0995-7146}\inst{\ref{aff83},\ref{aff46}}
\and G.~Sirri\orcid{0000-0003-2626-2853}\inst{\ref{aff19}}
\and J.~Skottfelt\orcid{0000-0003-1310-8283}\inst{\ref{aff89}}
\and L.~Stanco\orcid{0000-0002-9706-5104}\inst{\ref{aff46}}
\and P.~Tallada-Cresp\'{i}\orcid{0000-0002-1336-8328}\inst{\ref{aff90},\ref{aff35}}
\and H.~I.~Teplitz\orcid{0000-0002-7064-5424}\inst{\ref{aff91}}
\and I.~Tereno\inst{\ref{aff44},\ref{aff92}}
\and R.~Toledo-Moreo\orcid{0000-0002-2997-4859}\inst{\ref{aff93}}
\and F.~Torradeflot\orcid{0000-0003-1160-1517}\inst{\ref{aff35},\ref{aff90}}
\and I.~Tutusaus\orcid{0000-0002-3199-0399}\inst{\ref{aff94}}
\and L.~Valenziano\orcid{0000-0002-1170-0104}\inst{\ref{aff1},\ref{aff95}}
\and T.~Vassallo\orcid{0000-0001-6512-6358}\inst{\ref{aff22},\ref{aff49}}
\and A.~Veropalumbo\orcid{0000-0003-2387-1194}\inst{\ref{aff17},\ref{aff25}}
\and Y.~Wang\orcid{0000-0002-4749-2984}\inst{\ref{aff91}}
\and J.~Weller\orcid{0000-0002-8282-2010}\inst{\ref{aff22},\ref{aff15}}
\and A.~Zacchei\orcid{0000-0003-0396-1192}\inst{\ref{aff49},\ref{aff96}}
\and G.~Zamorani\orcid{0000-0002-2318-301X}\inst{\ref{aff1}}
\and J.~Zoubian\inst{\ref{aff61}}
\and E.~Zucca\orcid{0000-0002-5845-8132}\inst{\ref{aff1}}
\and M.~Bolzonella\orcid{0000-0003-3278-4607}\inst{\ref{aff1}}
\and C.~Burigana\orcid{0000-0002-3005-5796}\inst{\ref{aff97},\ref{aff95}}
\and P.~W.~Morris\orcid{0000-0002-5186-4381}\inst{\ref{aff98}}
\and V.~Scottez\inst{\ref{aff73},\ref{aff99}}
\and P.~Simon\inst{\ref{aff67}}
\and J.~Mart\'{i}n-Fleitas\orcid{0000-0002-8594-569X}\inst{\ref{aff100}}
\and D.~Scott\orcid{0000-0002-6878-9840}\inst{\ref{aff101}}}
										   
\institute{INAF-Osservatorio di Astrofisica e Scienza dello Spazio di Bologna, Via Piero Gobetti 93/3, 40129 Bologna, Italy\label{aff1}
\and
School of Mathematics and Physics, University of Surrey, Guildford, Surrey, GU2 7XH, UK\label{aff2}
\and
Leiden Observatory, Leiden University, Einsteinweg 55, 2333 CC Leiden, The Netherlands\label{aff3}
\and
Kapteyn Astronomical Institute, University of Groningen, PO Box 800, 9700 AV Groningen, The Netherlands\label{aff4}
\and
David A. Dunlap Department of Astronomy \& Astrophysics, University of Toronto, 50 St George Street, Toronto, Ontario M5S 3H4, Canada\label{aff5}
\and
Jodrell Bank Centre for Astrophysics, Department of Physics and Astronomy, University of Manchester, Oxford Road, Manchester M13 9PL, UK\label{aff6}
\and
Institute for Astronomy, University of Edinburgh, Royal Observatory, Blackford Hill, Edinburgh EH9 3HJ, UK\label{aff7}
\and
Department of Astrophysics/IMAPP, Radboud University, PO Box 9010, 6500 GL Nijmegen, The Netherlands\label{aff8}
\and
Observatoire Astronomique de Strasbourg (ObAS), Universit\'e de Strasbourg - CNRS, UMR 7550, Strasbourg, France\label{aff9}
\and
International Space University, 1 rue Jean-Dominique Cassini, 67400 Illkirch-Graffenstaden, France\label{aff10}
\and
National Astronomical Observatory of Japan, 2-21-1 Osawa, Mitaka, Tokyo 181-8588, Japan\label{aff11}
\and
Universit\'e de Strasbourg, CNRS, Observatoire astronomique de Strasbourg, UMR 7550, 67000 Strasbourg, France\label{aff12}
\and
AIM, CEA, CNRS, Universit\'{e} Paris-Saclay, Universit\'{e} de Paris, 91191 Gif-sur-Yvette, France\label{aff13}
\and
Max-Planck-Institut f\"ur Astronomie, K\"onigstuhl 17, 69117 Heidelberg, Germany\label{aff14}
\and
Max Planck Institute for Extraterrestrial Physics, Giessenbachstr. 1, 85748 Garching, Germany\label{aff15}
\and
ESAC/ESA, Camino Bajo del Castillo, s/n., Urb. Villafranca del Castillo, 28692 Villanueva de la Ca\~nada, Madrid, Spain\label{aff16}
\and
INAF-Osservatorio Astronomico di Brera, Via Brera 28, 20122 Milano, Italy\label{aff17}
\and
Dipartimento di Fisica e Astronomia, Universit\`a di Bologna, Via Gobetti 93/2, 40129 Bologna, Italy\label{aff18}
\and
INFN-Sezione di Bologna, Viale Berti Pichat 6/2, 40127 Bologna, Italy\label{aff19}
\and
INAF-Osservatorio Astronomico di Padova, Via dell'Osservatorio 5, 35122 Padova, Italy\label{aff20}
\and
Centre National d'Etudes Spatiales -- Centre spatial de Toulouse, 18 avenue Edouard Belin, 31401 Toulouse Cedex 9, France\label{aff21}
\and
Universit\"ats-Sternwarte M\"unchen, Fakult\"at f\"ur Physik, Ludwig-Maximilians-Universit\"at M\"unchen, Scheinerstrasse 1, 81679 M\"unchen, Germany\label{aff22}
\and
INAF-Osservatorio Astrofisico di Torino, Via Osservatorio 20, 10025 Pino Torinese (TO), Italy\label{aff23}
\and
Dipartimento di Fisica, Universit\`a di Genova, Via Dodecaneso 33, 16146, Genova, Italy\label{aff24}
\and
INFN-Sezione di Genova, Via Dodecaneso 33, 16146, Genova, Italy\label{aff25}
\and
Department of Physics "E. Pancini", University Federico II, Via Cinthia 6, 80126, Napoli, Italy\label{aff26}
\and
INAF-Osservatorio Astronomico di Capodimonte, Via Moiariello 16, 80131 Napoli, Italy\label{aff27}
\and
INFN section of Naples, Via Cinthia 6, 80126, Napoli, Italy\label{aff28}
\and
Instituto de Astrof\'isica e Ci\^encias do Espa\c{c}o, Universidade do Porto, CAUP, Rua das Estrelas, PT4150-762 Porto, Portugal\label{aff29}
\and
Dipartimento di Fisica, Universit\`a degli Studi di Torino, Via P. Giuria 1, 10125 Torino, Italy\label{aff30}
\and
INFN-Sezione di Torino, Via P. Giuria 1, 10125 Torino, Italy\label{aff31}
\and
Mullard Space Science Laboratory, University College London, Holmbury St Mary, Dorking, Surrey RH5 6NT, UK\label{aff32}
\and
INAF-IASF Milano, Via Alfonso Corti 12, 20133 Milano, Italy\label{aff33}
\and
Institut de F\'{i}sica d'Altes Energies (IFAE), The Barcelona Institute of Science and Technology, Campus UAB, 08193 Bellaterra (Barcelona), Spain\label{aff34}
\and
Port d'Informaci\'{o} Cient\'{i}fica, Campus UAB, C. Albareda s/n, 08193 Bellaterra (Barcelona), Spain\label{aff35}
\and
Institute for Theoretical Particle Physics and Cosmology (TTK), RWTH Aachen University, 52056 Aachen, Germany\label{aff36}
\and
INAF-Osservatorio Astronomico di Roma, Via Frascati 33, 00078 Monteporzio Catone, Italy\label{aff37}
\and
Dipartimento di Fisica e Astronomia "Augusto Righi" - Alma Mater Studiorum Universit\`a di Bologna, Viale Berti Pichat 6/2, 40127 Bologna, Italy\label{aff38}
\and
European Space Agency/ESRIN, Largo Galileo Galilei 1, 00044 Frascati, Roma, Italy\label{aff39}
\and
Universit\'e Claude Bernard Lyon 1, CNRS/IN2P3, IP2I Lyon, UMR 5822, Villeurbanne, F-69100, France\label{aff40}
\and
Institute of Physics, Laboratory of Astrophysics, Ecole Polytechnique F\'ed\'erale de Lausanne (EPFL), Observatoire de Sauverny, 1290 Versoix, Switzerland\label{aff41}
\and
UCB Lyon 1, CNRS/IN2P3, IUF, IP2I Lyon, 4 rue Enrico Fermi, 69622 Villeurbanne, France\label{aff42}
\and
Department of Astronomy, University of Geneva, ch. d'Ecogia 16, 1290 Versoix, Switzerland\label{aff43}
\and
Departamento de F\'isica, Faculdade de Ci\^encias, Universidade de Lisboa, Edif\'icio C8, Campo Grande, PT1749-016 Lisboa, Portugal\label{aff44}
\and
Instituto de Astrof\'isica e Ci\^encias do Espa\c{c}o, Faculdade de Ci\^encias, Universidade de Lisboa, Campo Grande, 1749-016 Lisboa, Portugal\label{aff45}
\and
INFN-Padova, Via Marzolo 8, 35131 Padova, Italy\label{aff46}
\and
INAF-Istituto di Astrofisica e Planetologia Spaziali, via del Fosso del Cavaliere, 100, 00100 Roma, Italy\label{aff47}
\and
Universit\'e Paris-Saclay, Universit\'e Paris Cit\'e, CEA, CNRS, AIM, 91191, Gif-sur-Yvette, France\label{aff48}
\and
INAF-Osservatorio Astronomico di Trieste, Via G. B. Tiepolo 11, 34143 Trieste, Italy\label{aff49}
\and
Istituto Nazionale di Fisica Nucleare, Sezione di Bologna, Via Irnerio 46, 40126 Bologna, Italy\label{aff50}
\and
Dipartimento di Fisica "Aldo Pontremoli", Universit\`a degli Studi di Milano, Via Celoria 16, 20133 Milano, Italy\label{aff51}
\and
INFN-Sezione di Milano, Via Celoria 16, 20133 Milano, Italy\label{aff52}
\and
Higgs Centre for Theoretical Physics, School of Physics and Astronomy, The University of Edinburgh, Edinburgh EH9 3FD, UK\label{aff53}
\and
Jet Propulsion Laboratory, California Institute of Technology, 4800 Oak Grove Drive, Pasadena, CA, 91109, USA\label{aff54}
\and
Department of Physics, Lancaster University, Lancaster, LA1 4YB, UK\label{aff55}
\and
von Hoerner \& Sulger GmbH, Schlossplatz 8, 68723 Schwetzingen, Germany\label{aff56}
\and
Technical University of Denmark, Elektrovej 327, 2800 Kgs. Lyngby, Denmark\label{aff57}
\and
Cosmic Dawn Center (DAWN), Denmark\label{aff58}
\and
Institut d'Astrophysique de Paris, UMR 7095, CNRS, and Sorbonne Universit\'e, 98 bis boulevard Arago, 75014 Paris, France\label{aff59}
\and
Department of Physics and Helsinki Institute of Physics, Gustaf H\"allstr\"omin katu 2, 00014 University of Helsinki, Finland\label{aff60}
\and
Aix-Marseille Universit\'e, CNRS/IN2P3, CPPM, Marseille, France\label{aff61}
\and
Universit\'e de Gen\`eve, D\'epartement de Physique Th\'eorique and Centre for Astroparticle Physics, 24 quai Ernest-Ansermet, CH-1211 Gen\`eve 4, Switzerland\label{aff62}
\and
Department of Physics, P.O. Box 64, 00014 University of Helsinki, Finland\label{aff63}
\and
Helsinki Institute of Physics, Gustaf H{\"a}llstr{\"o}min katu 2, University of Helsinki, Helsinki, Finland\label{aff64}
\and
Institute of Theoretical Astrophysics, University of Oslo, P.O. Box 1029 Blindern, 0315 Oslo, Norway\label{aff65}
\and
NOVA optical infrared instrumentation group at ASTRON, Oude Hoogeveensedijk 4, 7991PD, Dwingeloo, The Netherlands\label{aff66}
\and
Universit\"at Bonn, Argelander-Institut f\"ur Astronomie, Auf dem H\"ugel 71, 53121 Bonn, Germany\label{aff67}
\and
Aix-Marseille Universit\'e, CNRS, CNES, LAM, Marseille, France\label{aff68}
\and
Dipartimento di Fisica e Astronomia "Augusto Righi" - Alma Mater Studiorum Universit\`a di Bologna, via Piero Gobetti 93/2, 40129 Bologna, Italy\label{aff69}
\and
Department of Physics, Institute for Computational Cosmology, Durham University, South Road, DH1 3LE, UK\label{aff70}
\and
Universit\'e C\^{o}te d'Azur, Observatoire de la C\^{o}te d'Azur, CNRS, Laboratoire Lagrange, Bd de l'Observatoire, CS 34229, 06304 Nice cedex 4, France\label{aff71}
\and
Universit\'e Paris Cit\'e, CNRS, Astroparticule et Cosmologie, 75013 Paris, France\label{aff72}
\and
Institut d'Astrophysique de Paris, 98bis Boulevard Arago, 75014, Paris, France\label{aff73}
\and
European Space Agency/ESTEC, Keplerlaan 1, 2201 AZ Noordwijk, The Netherlands\label{aff74}
\and
Department of Physics and Astronomy, University of Aarhus, Ny Munkegade 120, DK-8000 Aarhus C, Denmark\label{aff75}
\and
Waterloo Centre for Astrophysics, University of Waterloo, Waterloo, Ontario N2L 3G1, Canada\label{aff76}
\and
Department of Physics and Astronomy, University of Waterloo, Waterloo, Ontario N2L 3G1, Canada\label{aff77}
\and
Perimeter Institute for Theoretical Physics, Waterloo, Ontario N2L 2Y5, Canada\label{aff78}
\and
Space Science Data Center, Italian Space Agency, via del Politecnico snc, 00133 Roma, Italy\label{aff79}
\and
Institute of Space Science, Str. Atomistilor, nr. 409 M\u{a}gurele, Ilfov, 077125, Romania\label{aff80}
\and
Instituto de Astrof\'isica de Canarias, Calle V\'ia L\'actea s/n, 38204, San Crist\'obal de La Laguna, Tenerife, Spain\label{aff81}
\and
Departamento de Astrof\'isica, Universidad de La Laguna, 38206, La Laguna, Tenerife, Spain\label{aff82}
\and
Dipartimento di Fisica e Astronomia "G. Galilei", Universit\`a di Padova, Via Marzolo 8, 35131 Padova, Italy\label{aff83}
\and
Departamento de F\'isica, FCFM, Universidad de Chile, Blanco Encalada 2008, Santiago, Chile\label{aff84}
\and
Universit\"at Innsbruck, Institut f\"ur Astro- und Teilchenphysik, Technikerstr. 25/8, 6020 Innsbruck, Austria\label{aff85}
\and
Institut d'Estudis Espacials de Catalunya (IEEC),  Edifici RDIT, Campus UPC, 08860 Castelldefels, Barcelona, Spain\label{aff86}
\and
Institute of Space Sciences (ICE, CSIC), Campus UAB, Carrer de Can Magrans, s/n, 08193 Barcelona, Spain\label{aff87}
\and
Satlantis, University Science Park, Sede Bld 48940, Leioa-Bilbao, Spain\label{aff88}
\and
Centre for Electronic Imaging, Open University, Walton Hall, Milton Keynes, MK7~6AA, UK\label{aff89}
\and
Centro de Investigaciones Energ\'eticas, Medioambientales y Tecnol\'ogicas (CIEMAT), Avenida Complutense 40, 28040 Madrid, Spain\label{aff90}
\and
Infrared Processing and Analysis Center, California Institute of Technology, Pasadena, CA 91125, USA\label{aff91}
\and
Instituto de Astrof\'isica e Ci\^encias do Espa\c{c}o, Faculdade de Ci\^encias, Universidade de Lisboa, Tapada da Ajuda, 1349-018 Lisboa, Portugal\label{aff92}
\and
Universidad Polit\'ecnica de Cartagena, Departamento de Electr\'onica y Tecnolog\'ia de Computadoras,  Plaza del Hospital 1, 30202 Cartagena, Spain\label{aff93}
\and
Institut de Recherche en Astrophysique et Plan\'etologie (IRAP), Universit\'e de Toulouse, CNRS, UPS, CNES, 14 Av. Edouard Belin, 31400 Toulouse, France\label{aff94}
\and
INFN-Bologna, Via Irnerio 46, 40126 Bologna, Italy\label{aff95}
\and
IFPU, Institute for Fundamental Physics of the Universe, via Beirut 2, 34151 Trieste, Italy\label{aff96}
\and
INAF, Istituto di Radioastronomia, Via Piero Gobetti 101, 40129 Bologna, Italy\label{aff97}
\and
California institute of Technology, 1200 E California Blvd, Pasadena, CA 91125, USA\label{aff98}
\and
Junia, EPA department, 41 Bd Vauban, 59800 Lille, France\label{aff99}
\and
Aurora Technology for European Space Agency (ESA), Camino bajo del Castillo, s/n, Urbanizacion Villafranca del Castillo, Villanueva de la Ca\~nada, 28692 Madrid, Spain\label{aff100}
\and
Department of Physics and Astronomy, University of British Columbia, Vancouver, BC V6T 1Z1, Canada\label{aff101}}          

 \date{May 14, 2024}

%
%
   \abstract{
As part of the \Euclid Early Release Observations (ERO) programme, we analyse deep, wide-field imaging from the VIS and NISP instruments of two Milky Way globular clusters (GCs), namely NGC~6254 (M10) and NGC~6397, to look for observational evidence of their dynamical interaction with the Milky Way. We search for such an interaction in the form of structural and morphological features in the clusters' outermost regions, which are suggestive of the development of tidal tails on scales larger than those sampled by the ERO programme. Our multi-band photometric analysis results in deep and well-behaved colour--magnitude diagrams that, in turn, enable an accurate membership selection. The surface brightness profiles built from these samples of member stars are the deepest ever obtained for these two Milky Way GCs, reaching down to $\sim30.0$ mag~arcsec$^{-2}$, which is about $1.5$ mag~arcsec$^{-2}$ below the current limit. The investigation of the two-dimensional density map of NGC~6254 reveals an elongated morphology of the cluster peripheries in the direction and with the amplitude predicted by $N$-body simulations of the cluster's dynamical evolution, at high statistical significance. We interpret this as strong evidence for the first detection of tidally induced morphological distortion around this cluster. The density map of NGC~6397 reveals a slightly elliptical morphology, in agreement with previous studies, which requires further investigation on larger scales to be properly interpreted. This ERO project thus demonstrates the power of \Euclid in studying the outer regions of GCs at an unprecedented level of detail, thanks to the combination of large field of view, high spatial resolution, and depth enabled by the telescope. Our results highlight the future \Euclid survey as the ideal data set to investigate GC tidal tails and stellar streams. }
%
%
%
%
   \titlerunning{\Euclid: ERO -- Milky Way globular clusters}
   \authorrunning{D. Massari et al.}
   
   \maketitle
%
%
%
%
   
\section{\label{sc:Intro}Introduction}
Globular clusters (GCs) have long played an important role as cosmic laboratories for a wealth of astronomical open questions. As the closest example in nature of simple stellar populations, they are exceptionally well suited to investigate stellar evolution \citep{salaris93, cassisi99}. Due to their short dynamical timescales, GCs are ideal for studying dynamical processes and their interplay with stellar evolution itself \citep{hut92, meylan97}. Moreover, being some of the oldest objects in the Universe, they are witnesses and powerful tracers of the early formation and evolution of galaxies like the Milky Way \citep{searle78, forbes10, massari19, kruijssen20, massari23}.

A major challenge when studying GCs arises from their richness in stars. Especially in the inner regions, their high stellar density causes crowding to be severe, so much so that these objects have been observed mostly with high-resolution, limited field-of-view (FoV) cameras, which have sufficient depth to sample their stellar component down to the faintest, lowest-mass stars. 
A clear example of this is provided by the success that the \HST  (HST) has demonstrated in the investigation of GCs \citep[e.g.,][]{sarajedini07, piotto15}, and in particular of their central regions. However, the information provided by low-mass stars in GC outskirts is crucial for understanding how these systems evolve both internally and in the presence of the Milky Way.

For example, it is well known that any stellar system embedded in a gravitational potential is subject to tidal forces \citep{binney87}. This effect is more pronounced where the tidal forces are comparable to, or start to dominate over the system's self-gravity, that is in the outer peripheries. Stars in the outskirts of GCs are thus more prone to have their orbits distorted by tidal forces, contributing to the development of a cluster tidal tail. Since the internal dynamical evolution of GCs drives faint, low-mass stars to preferentially populate the cluster outer regions, GC tidal tails appear as morphological features having very low surface brightness \citep{grillmair95, leon2000}.

The detection of tidal tails around GCs has proven challenging so far. On the one hand, dense GCs experience low evaporation rates during their evolution, so that stars in their tidal tails typically constitute less than 0.1\% of the surrounding stellar field \citep{dinescu99, balbinot18, sollima20}. On the other hand, tidal tails emerge over scales that can be several to tens of degrees wide \citep{odenkirchen01, grillmair06, erkal17}, making them difficult to sample with existing high-resolution imagers.
The advent of photometric surveys, such as SDSS \citep{sdss}, Gaia \citep{gaia}, and DES \citep{des}, has at least partially solved the latter of these issues, and led to the detection of tidal tails around several GCs \citep[e.g.,][]{odenkirchen01, belokurov06, nied10, sollima11, balbinot11, myeong17, navarrete17, ibata19, grillmair19, kaderali19, carballo19, bianchini19, Shipp_etal_2020,piatti21}. Yet, the number of systems displaying tidal tails remains small \citep{kuzma18, sollima20}. The absence of tails might be explained by the presence of dark matter halos surrounding GCs, which would prevent stars from escaping \citep{moore96}, and would favour the development of diffuse stellar envelopes \citep{penarrubia17, kuzma18}. Alternatively, the reason why tails remain such an elusive feature could be an observational bias, related to the fact that existing imagers are not able to observe very faint stars on large scales on the sky \citep{balbinot18}.

Thanks to its wide-field imaging capabilities, the \Euclid mission \citep{Laureijs11, EuclidSkyOverview} will provide a definitive answer to this fundamental open question concerning the nature of GCs. The Euclid Wide Survey will observe an area of approximately $14\,000$ deg$^2$ with exceptionally deep AB limiting magnitudes (5$\sigma$ point-like sources) of $26.2$ in the optical VIS band (\IE) and $24.5$ in the near-infrared NISP bands \citep[\YE, \JE, \HE, see][]{Scaramella-EP1}. This means that all of the GCs covered by the survey footprint will be observed with an unprecedented combination of depth, spatial coverage, and resolution, in four different photometric bands.

In this paper, we showcase the power of \Euclid for the investigation of Milky Way GC peripheries by analysing Early Release Observations \citep[ERO,][]{EROcite} of two GCs, namely NGC~6254 (M10) and NGC~6397.
NGC~6254 has already been the subject of some wide-field photometric studies. In the first of these, \cite{leon2000} analysed photometric plates taken with the ESO Schmidt telescope, covering a $5.5\times5.5$ deg$^2$ area around the cluster and having a limiting magnitude of $R\simeq19$, thus barely reaching the cluster main-sequence turn-off. The authors found the presence of tidal tails extending along the north--south direction, but they highlighted a possible bias in the detection induced by the strong gradient in the dust extinction across the field.
In a later study, \cite{dalessandro13} investigated the density profile of the cluster using deeper ($V\simeq20$) photometry obtained from observations taken with the Wide-Field Imager mounted at the 2.2-m ESO-MPG telescope over a $33\times34$ arcmin$^{2}$ FoV. In this case, the authors found a density profile that was well fit by a simple King model \citep{king62}, thus possibly indicating a lack of asymmetric morphological features like tidal tails.
According to the predictive algorithm by \cite{balbinot18}, which takes into account the mass loss experienced by a GC and its orbital phase, NGC~6254 has a high chance of having tidal tails as it is very close to its apocentre (orbital phase $\Phi=0.91$) and it should have already lost 61\% of its initial mass.

The detection of tidal tails around NGC~6397 is debated, too. The first to look for extra-tidal features were \cite{leon2000}, who found overdensities that were classified as unreliable, due to the uncertainty in the distribution of dust around the cluster. More recently, as Gaia-related results, \cite{ibata21, ibata23} found the presence of a possible tail extending more than 18 degrees on the sky, but \cite{boldrini21} challenged these findings by not detecting any tail. According to the metric by \cite{balbinot18}, NGC~6397 is a strong candidate for having developed tidal tails, as it has lost 72\% of its initial mass and its orbital phase is $\Phi=0.95$.

As part of the \Euclid ERO programme, the available imaging covers an almost square region of about $0.8 \times 0.8$ deg$^{2}$ centred on the GCs. At a distance of $5.07$ kpc \citep{baumgardt21}, the tidal radius of NGC~6254 is about $r_{\rm t}\sim0.3$ deg \citep{dalessandro13}, and thus will be entirely covered by the observations. The spatial coverage is more limited for NGC~6397, whose tidal radius is larger, being $r_{\rm t}\sim0.6$ deg \citep{moreno14}, at a distance of $2.48$ kpc \citep{baumgardt21}. This means that it is likely these ERO data will not enable the detection of tidal tails, which start to develop well outside the tidal radius, at the location of the Jacobi radius \citep[$0.6$ deg for NGC~6254 and $1.15$ deg for NGC~6397, see][]{webb13}. Nonetheless, they will give us an unprecedented chance to detect the morphological distortions in the clusters' outer regions, which may ultimately hint at the presence of tidal tails on larger scales.

This paper is structured as follows. In Sect.\,\ref{sec:data}, the observations of the two GCs are described. In Sect.\,\ref{sec:photo}, we present the results of the photometric analysis by showing the first \Euclid colour--magnitude diagrams (CMDs, the magnitudes are in the AB photometric system) of Milky Way GCs. The morphology of the GCs' outer regions is discussed in Sect.\,\ref{sec:results}, and it is further interpreted in light of $N$-body simulations in Sect.\,\ref{sec:nbody}. Finally, the conclusions of our work are summarised in Sect.\,\ref{sec:concl}.

\section{Early Release Observations of NGC~6254 and NGC~6397}\label{sec:data}

\begin{figure*}[tbp!]
\centering
\includegraphics[width=\textwidth]{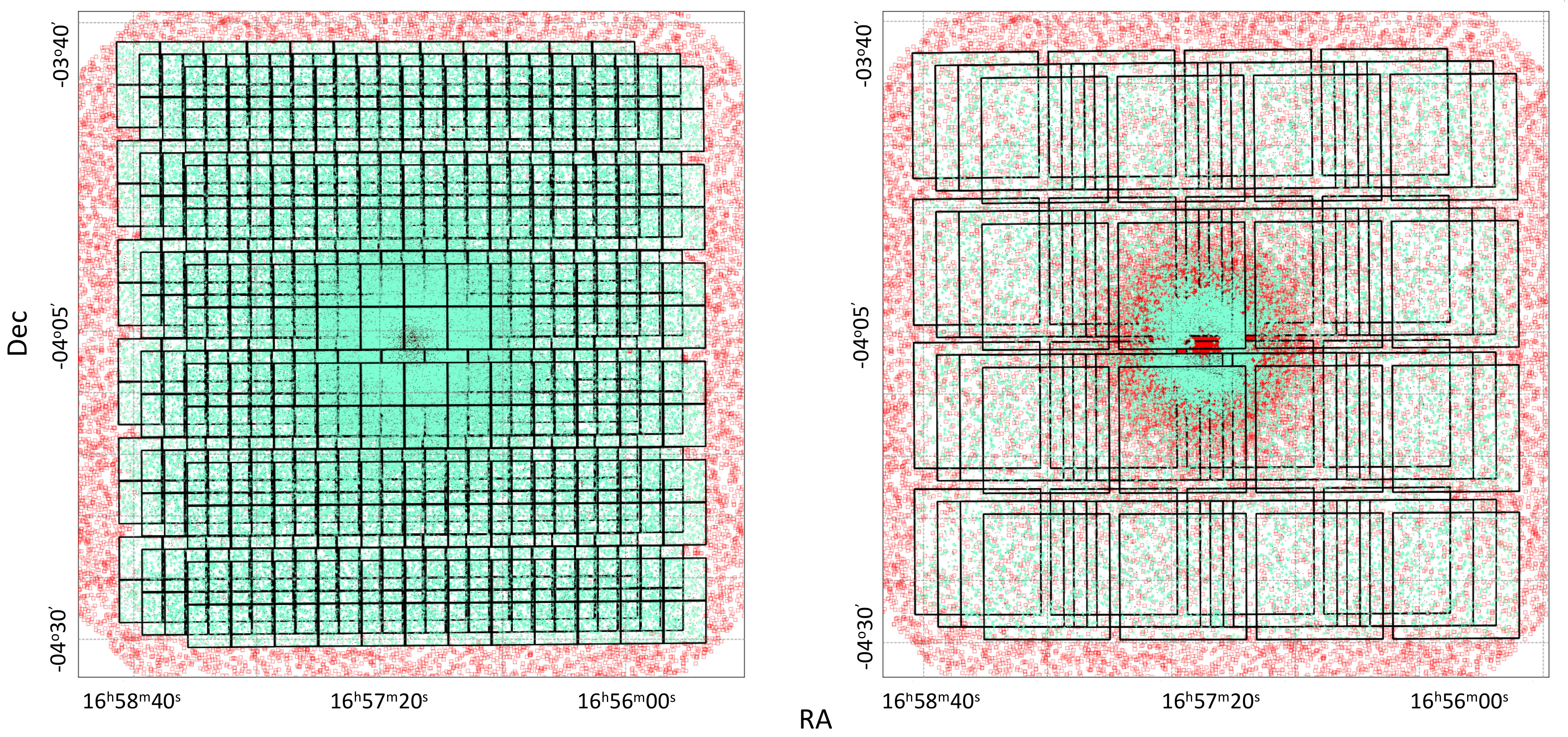}
\caption{Sky coverage and dither pattern of the four VIS (left-hand panel) and NISP (right-hand panel) exposures of the ERO ROS for NGC~6254. Red symbols indicate {\it Gaia} reference stars. Green symbols represent \Euclid detections.}
\label{fig:vis_stack}
\end{figure*}

The observations of NGC\,6254 and NGC\,6397 have been performed on the 9th and the 22nd of September 2023, respectively, as part of the \Euclid ERO Programme. This programme \citep[which is described in detail in][C24 hereafter]{EROData} further includes observations of the $\sigma$\,Orionis cluster \citep[][]{EROOrion}, of nearby galaxies \citep[][]{ERONearbyGals}, of the Fornax galaxy cluster \citep[][]{EROFornaxGCs}, of the Perseus cluster of galaxies \citep[][]{EROPerseusOverview, EROPerseusICL, EROPerseusDGs} and of a giant gravitational lens \citep[][]{EROLensData, EROLensVISDropouts}. Each observation analysed here consists of a single reference observation sequence \citep[ROS, see][]{Scaramella-EP1} centred on the cluster. One ROS includes four different dithered pointings, each resulting in one $560$ seconds-long exposure in the \IE band  \citep{EuclidSkyVIS}, and in $87.2$ seconds-long exposures in the \YE, \JE and \HE bands\footnote{The quoted numbers refer to the effective exposure times, whereas the total duration indicated in \cite{Scaramella-EP1} includes overheads.}, plus one spectral exposure that is neglected for this study. All of the exposures have been reduced by means of the data reduction pipeline described by C24. In the following, we briefly describe the main outputs of the pipeline.

\subsection{VIS and NISP imaging}

Figure~\ref{fig:vis_stack} shows the distribution on the plane of the sky of the four deep photometric exposures of NGC~6254 ROS in the \IE and NISP bands.
The VIS exposures have been stacked together following the prescriptions in C24, achieving an internal 1$\sigma$ astrometric precision of $\sim4$ mas. The same stacking procedures of C24 have been applied on the NISP \citep{Schirmer-EP18, EuclidSkyNISP} exposures as well.
In this case the astrometric referencing onto Gaia DR3 positions achieved a 1$\sigma$ precision of 15\,mas, which is still less than one tenth of the pixel scale \cite[0.3\,arcsec\,pixel$^{-1}$, see][]{Schirmer-EP18}. 

The measured full width at half maximum (FWHM) in the stacked images is typically $0.16$\,arcsec in the VIS band and $\sim0.4$ arcsec in the NISP ones, and is stable across the FoV within a few percent. 

\begin{figure*}[htbp!]
\centering
\includegraphics[width=\textwidth]{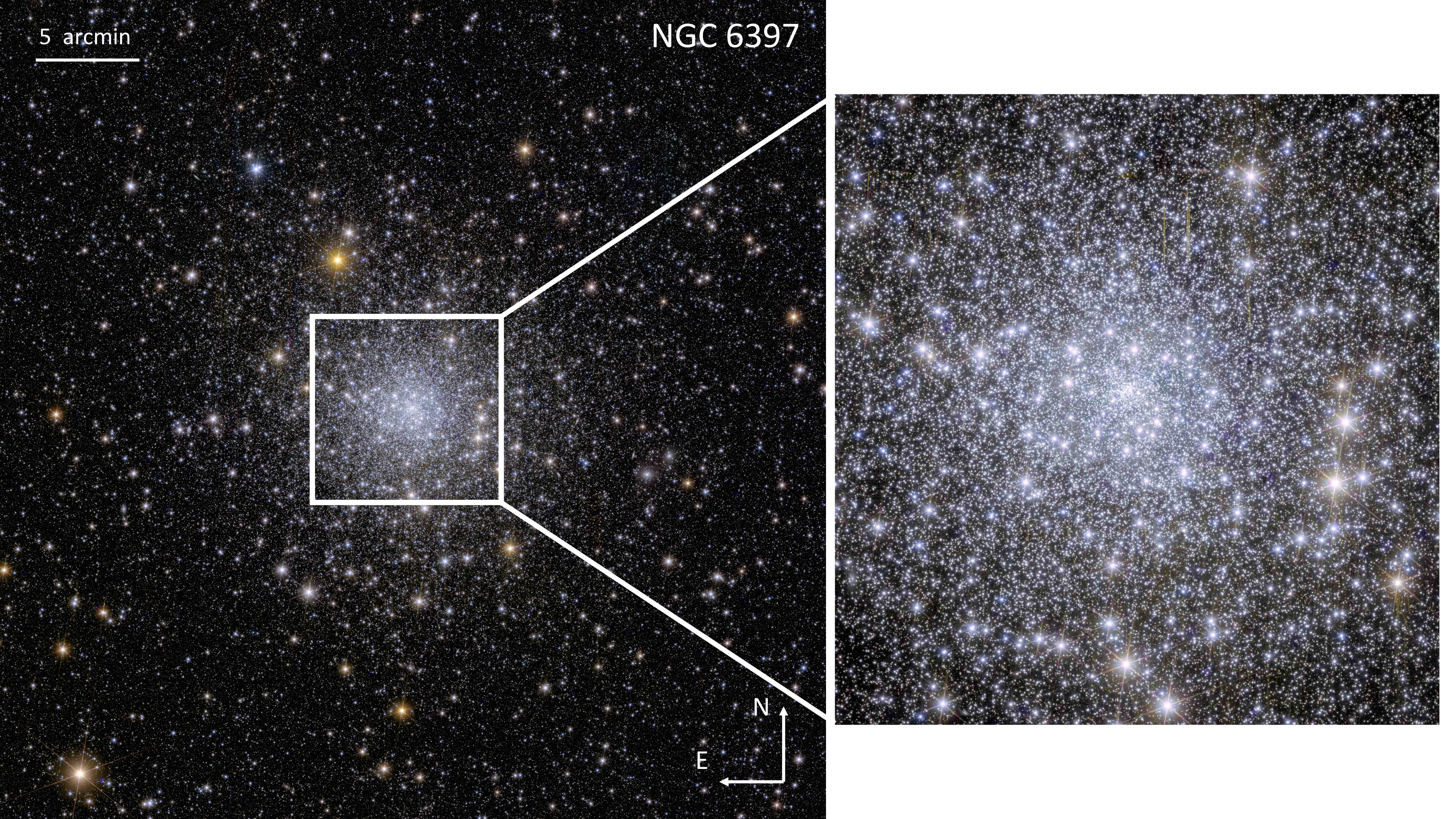}
\caption{RGB image of NGC~6397. The left-hand panel shows the entire FoV covered by the observations. The right-hand panel shows a $10\times10$\,arcmin$^2$ zoomed-in detail in the GC core. Credits: ESA/\Euclid/Euclid Consortium/NASA, image processing by J.-C. Cuillandre (CEA Paris-Saclay), G. Anselmi, CC BY-SA 3.0 IGO.}
\label{fig:vis_image}
\end{figure*}

The dither pattern allowed for a continuous coverage of the field around the clusters. The large FoV coverage and high spatial resolution are exceptionally represented by, respectively, the left- and the right-hand panels of Fig.~\ref{fig:vis_image}, which show an RGB image of the field around NGC~6397.
The FoV coverage of NGC~6254 resulted in a slightly imperfect gap filling due to the dither pattern. This does not impact our scientific objectives.

\section{Photometric analysis}\label{sec:photo}

The stacked images described above have been analysed using the {\tt Astromatic} software suite, including {\tt Sextractor} and {\tt PSFEx} \citep{psfex}.
Briefly, the {\tt PSFEx} code models the point spread function (PSF) of an instrument as a combination of basis functions and works directly on the image. In particular, {\tt PSFEx} selects point-like sources detected by {\tt Sextractor}, and uses them to compute the PSF model. For both the GCs investigated in this work, we selected stars for the PSF model that are bright, non-saturated, and relatively isolated in each band; specifically, we only selected point-like sources (with a maximum ellipticity of 0.3 and a FWHM within 30\% of the nominal one) having at least $5$ pixels above a detection threshold of 10\,$\sigma$ above the background. A further filtering through a Moffat function has been imposed to reject spurious detections. The PSF model has been determined over a $6\times6$ equally spaced grid covering the whole FoV, allowing for a linear spatial variation.
In this way, for both the GCs, the number of sources used to model the PSF exceeded a few thousand. The final PSF models have an average ellipticity of $0.02$ and a FWHM spatial variation of only $3\%$.

These models have then been used by {\tt Sextractor} to perform PSF photometry on a much larger sample of sources, defined by relaxing the source-detection threshold to having even a single pixel 1.5\,$\sigma$ above the background.
In this way, we PSF-fit more than $400\,000$ sources in the VIS image of NGC~6254, and more than one million sources in the VIS image of NGC~6397. The number of detections in the NISP images is obviously lower because of reduced depth and resolution, about $190\,000$ for NGC~6254 and about $350\,000$ for NGC~6397. 
By combining the VIS and NISP catalogues, most of the artefacts and spurious detections (e.g., PSF spikes around very bright stars) are rejected, and the final multi-band catalogues for NGC~6254 and NGC~6397 contain positions, magnitudes, related uncertainties, and shape/quality parameters for $133\,090$ and $296\,646$ sources, respectively. Aperture corrections of a few percent \citep[consistent with other works on ERO data, see e.g.,][]{EROFornaxGCs}, estimated using bright and isolated stars, are applied to the measured magnitudes. Because of the very high crowding, only a modest number of stars have been fit in the core of the clusters, where incompleteness is therefore very high. We remark, though, that throughout this paper we do not focus on the clusters' central regions, and thus our conclusions are not affected by this. A quantitative assessment of the local completeness achieved by the ERO data of NGC~6254 in the VIS band is provided in the next section.


\subsection{Completeness: local estimate for NGC~6254}

Directly estimating the completeness of the photometry would require artificial-star tests that are computationally very expensive for images this large in size, and that are not yet available in the adopted data-reduction pipeline.
Nonetheless, we can take advantage of the particularly well suited, publicly available photometric catalogue by \cite{dalessandro11}. These authors analysed {HST} Wide Field Planetary Camera 2 (WFPC2) observations in the F606W and F814W bands of a stellar field located between 1 and 2~$r_{\rm h}$ of the cluster centre \citep[$r_{\rm h}=2.02$ arcmin according to][]{baumgardt21}. By means of artificial-star tests performed in \cite{beccari10}, the authors computed the completeness of their WFPC2 photometric catalogue, which we can therefore use for a relative comparison.

We cross-matched our VIS catalogue with the public photometry of \cite{dalessandro11} by means of a linear transformation, requiring matching sources to (a) be located within a distance of 0.5 arcsec and (b) have a difference between the \Euclid $\IE$ and the WFPC2 $m_{\rm F814W}$ magnitudes smaller than a conservative value of $2$ mag. The residuals from the resulting linear transformation solution are only $0.03$\,arcsec in both Right Ascension (RA) and Declination (Dec), thus ensuring an optimally performed cross-match.

\begin{figure}[htbp!]
\centering
\includegraphics[width=\columnwidth]{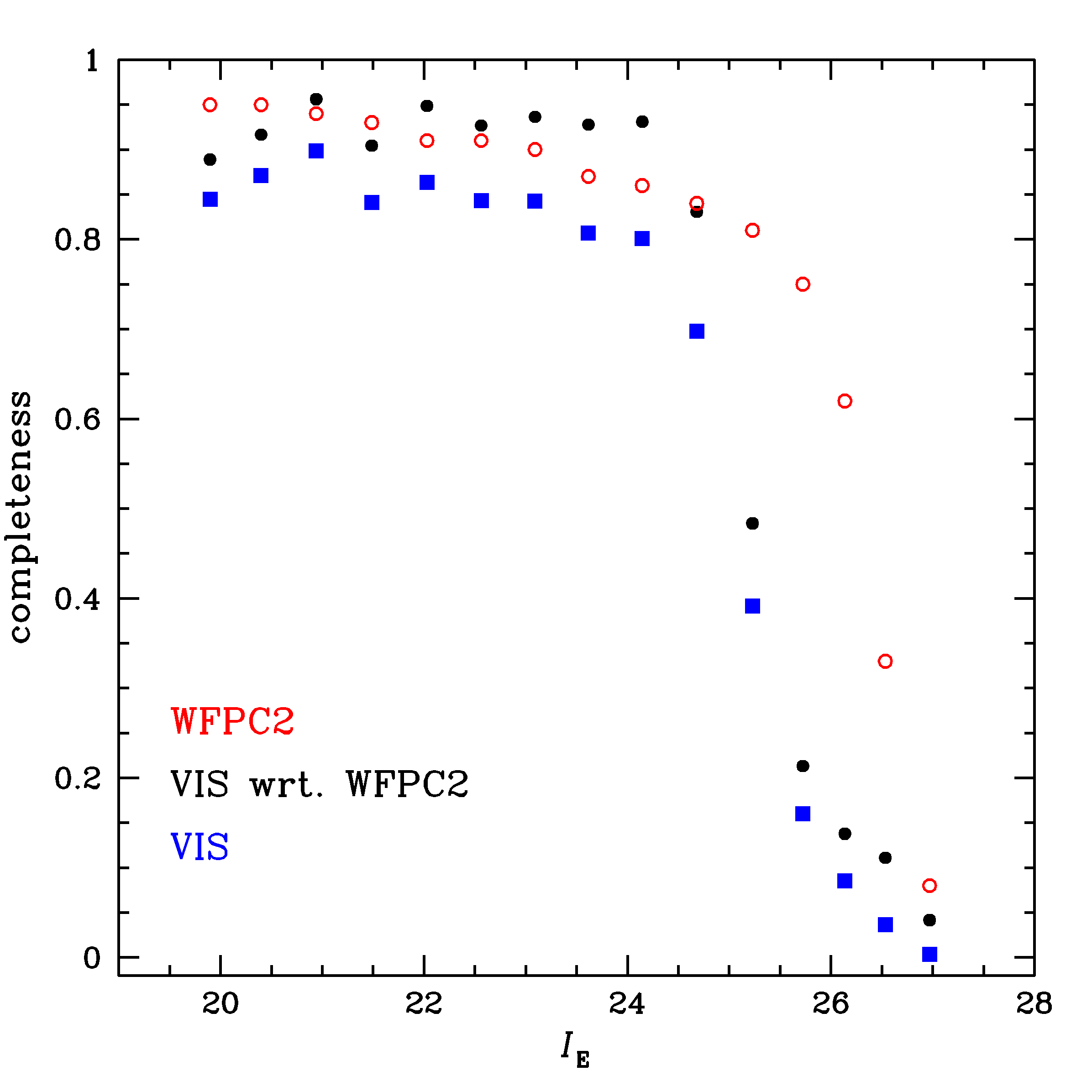}
\caption{Completeness of the \Euclid ERO in the $\IE$ band, at a distance of 1.5$\,r_{\rm h}$ from the core of NGC~6254. Red empty circles indicate the completeness of the reference {HST}/WFPC2 catalogue. Black circles represent the relative completeness of \Euclid with respect to {HST}. Finally, the filled blue squares show the absolute completeness of the VIS data.}
\label{fig:compl}
\end{figure}

Figure~\ref{fig:compl} shows the results of the comparison. We remark that, in order to perform a meaningful analysis, we decided to exclude regions around very bright, saturated stars, which are populated by artefacts and PSF structures in both the \Euclid and {HST} images.
The red open symbols in Fig.~\ref{fig:compl} show the absolute completeness as a function of magnitude for the WFPC2 sample, as estimated by \cite{dalessandro11}. The $m_{\rm F814W}$ magnitudes were converted to $\IE$ magnitudes by means of a linear transformation depending on $m_{\rm F555W}-m_{\rm F814W}$ colour, determined by using the stars in common between the two catalogues.
On the other hand, the black filled symbols indicate the {\it relative} completeness of the VIS catalogue with respect to the WFPC2 one. As shown in the Figure, the trend remains near-constant around a value of $94\%$ down to $\IE=24$, and then drops to lower percentages.
The product of these two curves, shown as blue filled squares in Fig.~\ref{fig:compl}, provides an estimate for the {\it absolute} completeness of the VIS catalogue, at an average distance from the GC centre of 1.5 $r_{\rm h}$. This is significantly closer to the most crowded regions of the GC than the peripheries we are interested in. Moreover, this estimate does not take into account possible VIS detections that are missing in the WFPC2 catalogue. For these reasons, it is safe to consider this estimate of the catalogue completeness as a {\it lower limit}. Completeness is certainly better at larger distances in our data.
As can be seen, the completeness remains about constant at  $\sim82\%$ down to $\IE=24$, dropping to $50\%$ completeness by $\IE\simeq25.0$.

The analysis of the VIS catalogue completeness has been performed to understand the ERO data better and to quantify the performance achieved, but it is not used for the study presented in this paper. These findings are of fundamental importance for future analysis of these ERO data, though, as a wealth of science cases, such as the investigation of the cluster mass functions, or of their binary populations, requires the knowledge of the photometric completeness and its correction.

\subsection{Colour--magnitude diagrams}

A fundamental quality check on the photometric analysis of GC fields comes from the cluster colour--magnitude diagram (CMD). In the following, we describe the most important features we found in the CMDs of NGC~6254 and NGC~6397, which are the first \Euclid CMDs of Milky Way stellar systems ever presented.

\subsubsection{NGC~6254}
\begin{figure}[htbp!]
\centering
\includegraphics[width=\columnwidth]{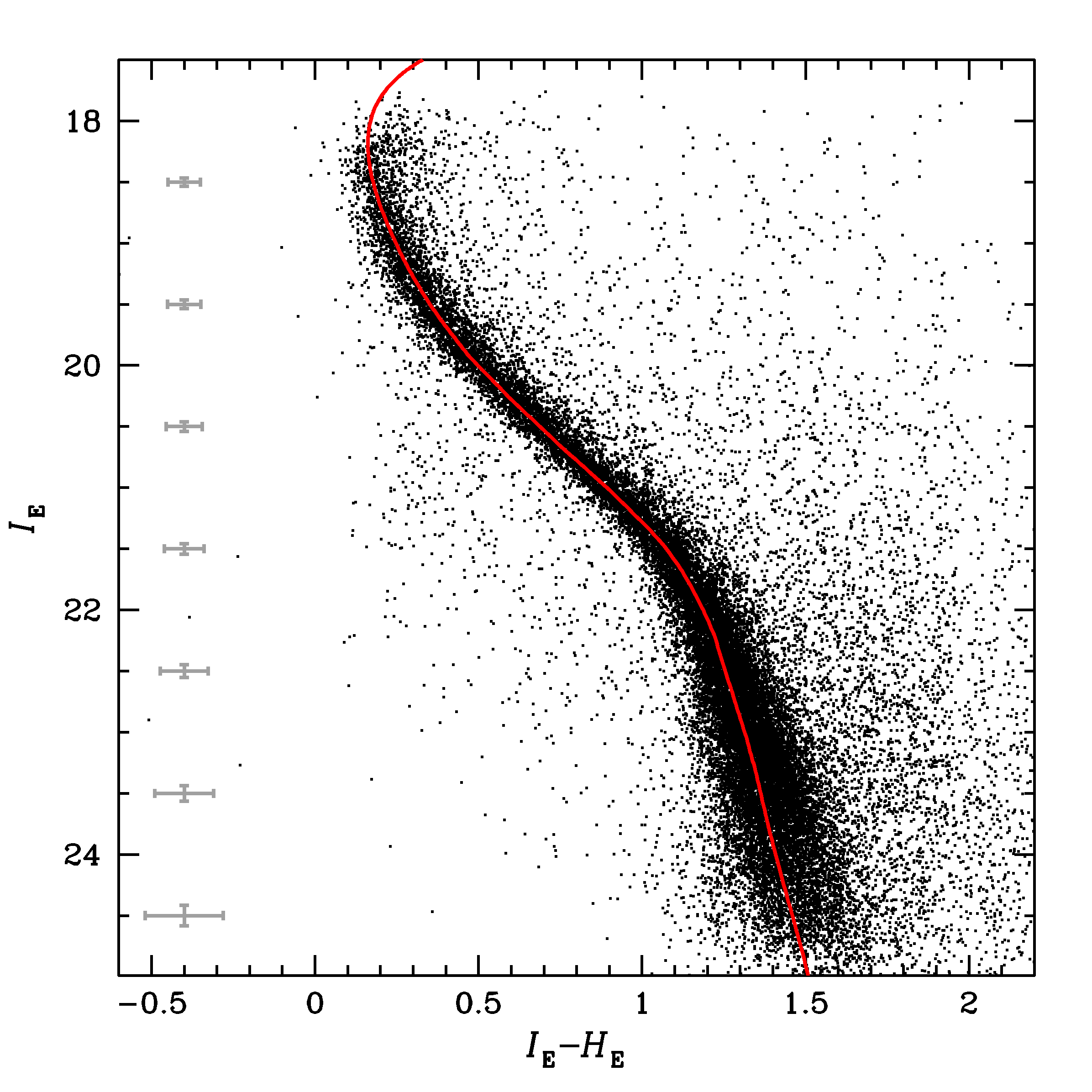}
\caption{Differential reddening corrected, ($\IE-\HE$, $\HE$) CMD for an annulus between $300$ arcsec and $600$ arcsec from the centre of NGC~6254. Error bars are marked in grey. A 13 Gyr old, ${\rm [Fe/H]}=-1.5$ dex and $\alpha$-enhanced theoretical isochrone from the BaSTI archive is shown in red for the sake of comparison.}
\label{fig:cmd}
\end{figure}

\begin{figure}[htbp!]
\centering
\includegraphics[width=\columnwidth]{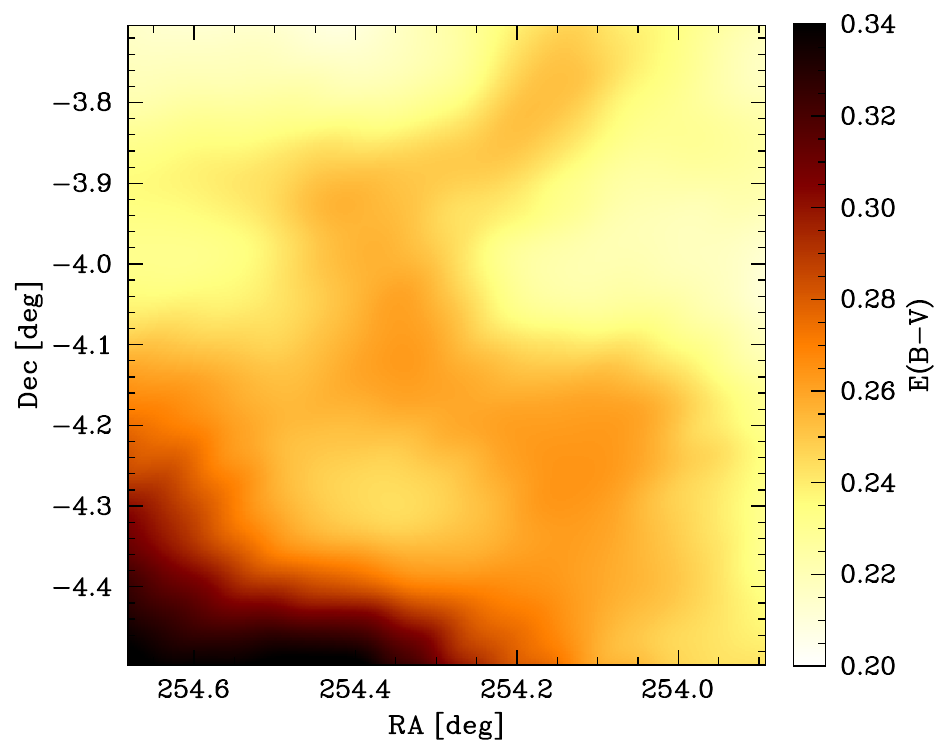}
\caption{Extinction map around NGC~6254 as obtained from the \cite{schlafly11} corrections to the extinction map by \cite{schlegel98}.}
\label{fig:extinction}
\end{figure}

Figure~\ref{fig:cmd} shows the ($\IE-\HE$, $\IE$) CMD for an annulus between $5$ and $10$ arcmin from the cluster centre, as determined by \cite{dalessandro13}. This region has been selected as it is close enough to the cluster for the CMD to be dominated by GC stars, but far enough (well beyond $2\,r_{\rm h}$) to avoid severe incompleteness and crowding effects. The photometry has been corrected for differential reddening by using the colour excess $E(B-V)$ map from \citet{schlegel98} evaluated using \texttt{mwdust} \citep{Bovy_2016} with the corrections from \citet{schlafly11} and the extinction law by \cite{cardelli89}. We correct all magnitudes to a fiducial value of $E(B-V)=0.28$, the mean absolute colour excess of the cluster \citep[][see Fig.~\ref{fig:extinction}]{harris96}. To improve upon the correction provided by the quoted map, we further apply the differential reddening correction procedure described by \cite{milone12}. By design, such an additional step effectively corrects also for local residuals in the photometric calibration, which can manifest as slight systematic differences in colour. Indeed, we find additional corrections of the order of a few hundredth mag, which we have applied to the observed magnitudes in Fig.~\ref{fig:cmd}.
Overall, the CMD shows a well-defined main sequence (MS), which extends from the turn-off (MSTO) point at $\IE\approx 18.5$ mag (where the deep exposures start to saturate) to a couple of magnitudes below the MS knee \citep[see e.g.][]{bono10, massari16, saracino18}, located at about $\IE=21.5$ mag.

A direct comparison with a theoretical isochrone is also shown. The isochrone has been taken from the BaSTI archive \citep{pietrinferni21} and describes a 13 Gyr old population having ${\rm [Fe/H]}=-1.5$ dex and $\alpha$-enhanced ([$\alpha$/{\rm Fe}]\,=\,0.4) chemical composition \citep{haynes08}.
To take into account possible uncertainties on the absolute photometric calibration, as well as on the adopted reddening estimates \citep{harris96}, the isochrone has been shifted in colour by a small arbitrary amount ($\delta[\IE-\HE]=0.02$) to provide a satisfactory fit. Once the shift is applied, the model describes the observed MS very well and shows that the observed CMD samples masses down to $0.16$ M$_{\odot}$ (at $\IE\simeq24.9$). This is the deepest CMD of a GC over such a large scale.

Another interesting feature that can be appreciated from this CMD is the possible widening of the faint MS below the MS knee. From a statistical point of view, the observed colour dispersion of upper-MS stars in the bright, non-saturated magnitude range $18.75<\IE<19.25$ is $\sigma_{\rm obs}^{\rm uMS}=0.04$. The typical nominal colour error at those magnitudes, as estimated by {\tt Sextractor}, is of only $0.01$ mag. Given that NGC~6254 is not a peculiar type-2 GC \citep{milone19}, and therefore should not present photometric features related to large iron or helium dispersion, which could significantly widen the observed sequences, such a nominal error seems to underestimate the actual one. We thus take the observed standard deviation $\sigma_{\rm obs}^{\rm uMS}$ of GC members at these bright magnitudes as a more robust (even though conservative) estimate of the photometric colour uncertainty. Possible terms that could inflate this value, making it an overestimate of the actual error, are the presence of photometric binaries, small helium variations among cluster stars, and residuals in the differential reddening correction. The observed binary fraction of NGC~6254 at the distance sampled by the CMD shown in Fig.~\ref{fig:cmd} is of only 1.5\% \citep{dalessandro11}, and therefore should not contribute significantly to $\sigma_{\rm obs}^{\rm uMS}$. According to \cite{milone18}, NGC~6254 might be characterised by a small helium spread $\Delta Y<0.029$, that could contribute to a photometric spread of 0.01--0.02\,mag.

This colour dispersion should also increase for fainter stars due to the decrease in the signal-to-noise ratio (SNR), according to $\sigma_{\rm obs}^{\rm uMS}\simeq\logten(1+1/{\rm SNR})$. Assuming a systematic colour uncertainty floor of $\sigma_{\rm obs}^{\rm uMS}=0.04$ at $\IE=19$, we thus expect $\sigma_{\rm obs}^{\rm uMS}=0.09$ below the MS knee, in the range $23<\IE<23.5$ (see the grey error bars in Fig.~\ref{fig:cmd}, where the error in magnitude has been computed by assuming an equal contribution to $\sigma_{\rm obs}$ from the two bands). However, the observed colour dispersion at the same magnitudes is somewhat larger, reaching $\sigma_{\rm obs}=0.12$. Given our conservative colour error estimate, we tentatively interpret this widening as an intrinsic feature, and cautiously attribute it to the GC's multiple stellar populations \citep[see e.g.,][]{gratton04, milone17}, which manifest in near-IR CMDs as a widening/splitting of the MS below the MS-knee due to the opacity effect of collision-induced absorption by water on the surface of M-dwarfs \citep{milone12, milone19}. The presence of multiple-populations in NGC~6254 is widely known, both from photometric \citep{monelli2013, milone17} and spectroscopic \citep{carretta09} studies. Yet, future more detailed analysis of this CMD will test the significance of this feature on more solid statistical ground, thus assessing the effectiveness of \Euclid in investigating this peculiar property of GCs.

\subsubsection{NGC~6397}

\begin{figure}[htbp!]
\centering
\includegraphics[width=\columnwidth]{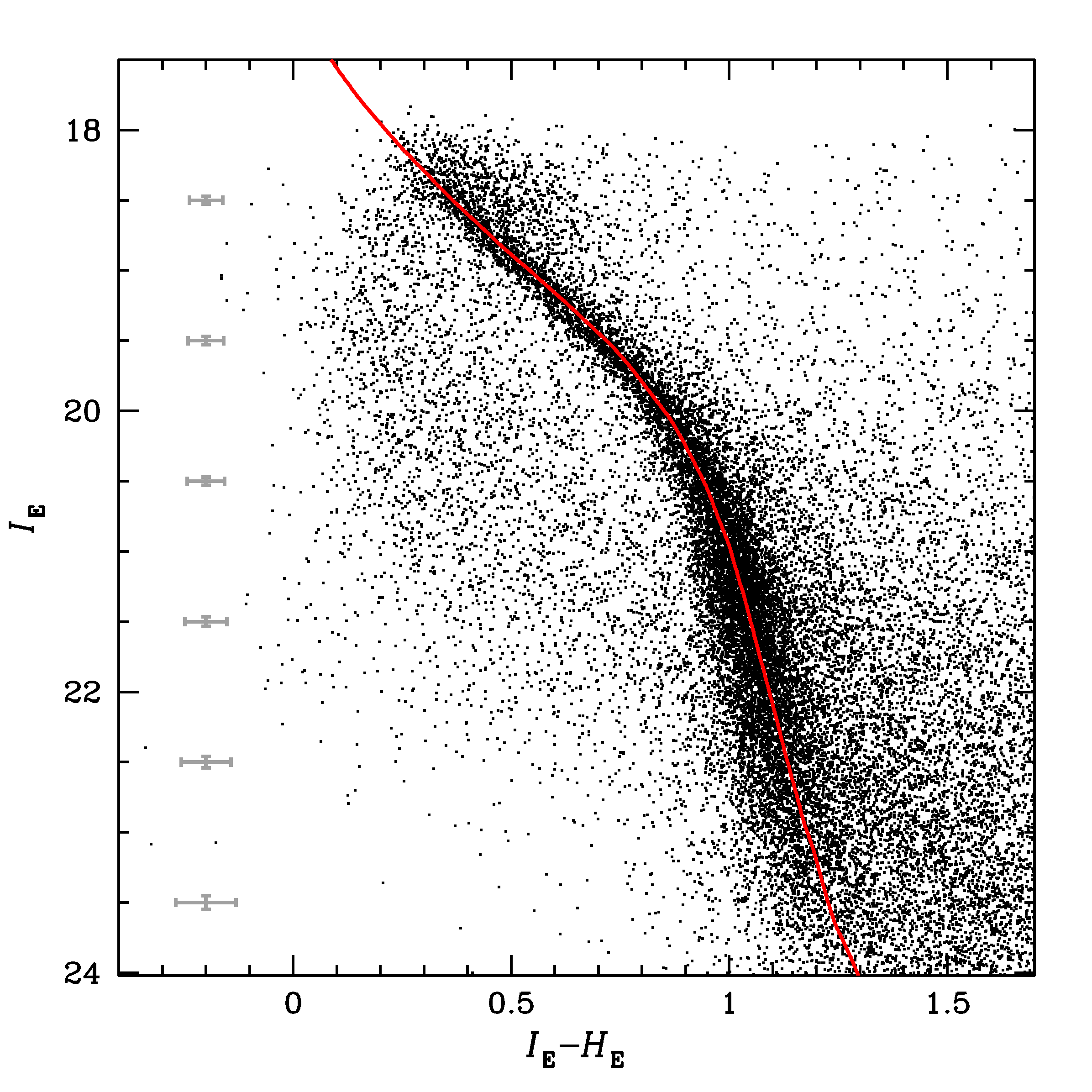}
\caption{Differential reddening corrected, ($\IE-\HE$, $\IE$) CMD for an annulus between 300 arcsec and 600 arcsec from the centre of NGC~6397. Error bars are marked in grey. A 13 Gyr old, ${\rm [Fe/H]}=-2.0$ dex, and $\alpha$-enhanced theoretical isochrone from the BaSTI archive is shown in red for the sake of comparison. The effect of saturation becomes evident at $\IE\lesssim18.5$}
\label{fig:cmd6397}
\end{figure}

Figure~\ref{fig:cmd6397} shows the ($\IE-\HE$, $\IE$) CMD for an annulus at distances $r\in[300,600]\,{\rm arcsec}$ from the centre of NGC~6397. The CMD is corrected for differential reddening and residuals on the photometric calibration as described for NGC~6254.

Also for NGC~6397, the cluster MS stands out very clearly as a well-defined and populated sequence, even though the contamination by field stars and background galaxies appears stronger than in the case of NGC~6254. The higher density of sources is likely the cause for the shallower depth reached by the CMD. The faintest end of the cluster MS seems to reach a magnitude of $\IE\sim24.0$, about $1$ mag brighter than in the case of NGC~6254. Such a shallower cutoff is imposed by the limiting magnitudes in the \HE band detections, that suffer from the low spatial resolution in such a crowded field, and from persistence effects.

Because of the closer distance of NGC~6397 to us, saturation prevents photometric measurement of the MSTO, but we are able to sample even lower stellar masses than in the case of NGC~6254. According to the BaSTI isochrone overplotted in Fig.~\ref{fig:cmd6397}, which represents a 13 Gyr old population with ${\rm [Fe/H]}=-2.0$\,dex and $\alpha$-enhanced chemical composition, the MS sampled lower limit corresponds to $\sim0.12$ M$_{\odot}$. 

The colour dispersion around the non-saturated upper MS (at $\IE\lesssim18.5$ mag) gives once again a reasonable upper limit to the achieved photometric error of $\sigma_{\rm obs}^{\rm uMS}=0.04$ mag, very similar to that found for NGC~6254. Following the same argument as in that cluster, the colour error below the MS knee, at $22<\IE<22.5$ should increase to $0.06$ for NGC~6397 (see the error bars in Fig.~\ref{fig:cmd6397}). Again, we detect a marginal widening of the faint MS ($\sigma_{\rm obs}=0.08$), possibly ascribable to the already known phenomenon of multiple populations in this cluster \citep[see e.g.,][]{dicriscienzo10, milone18, carretta09}.

\subsection{Star-galaxy and cluster-field separation}

\begin{figure}[htbp!]
\centering
\includegraphics[width=\columnwidth]{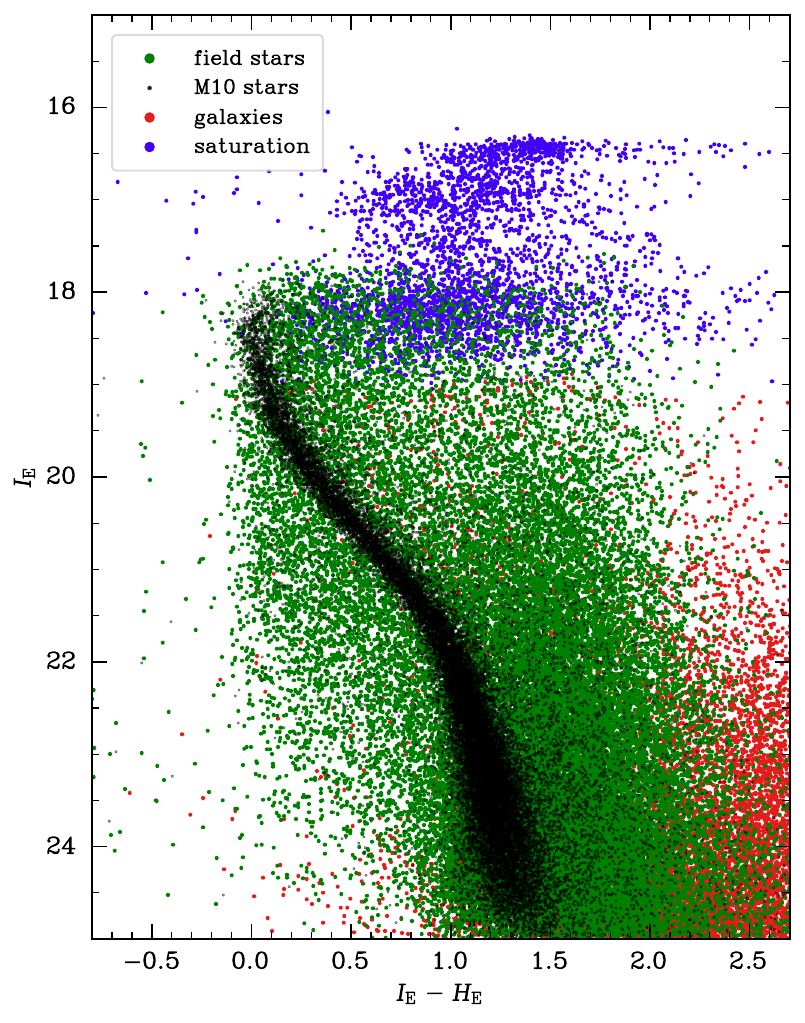}
\caption{CMD showing likely members of NGC~6254 (black), field stars (green), and background galaxies (red).}
\label{fig:cmd_types}
\end{figure}

\begin{figure}[htbp!]
\centering
\includegraphics[width=\columnwidth]{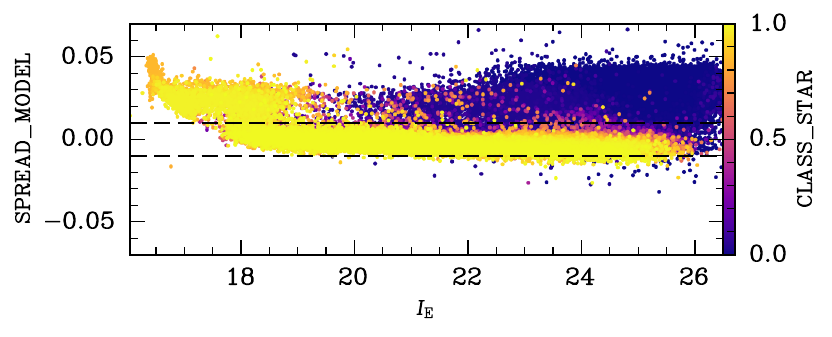}
\caption{\texttt{SPREAD\_MODEL} as a function of $\IE$ colour-coded by \texttt{CLASS\_STAR}. The dashed lines indicate a selection of $\lvert$\texttt{SPREAD\_MODEL}$\rvert<0.01$.}
\label{fig:sg}
\end{figure}

In order to study the morphology of the outer regions of the two GCs, it is fundamental to separate likely member GC stars from contaminating background galaxies and Milky Way disc stars. In the absence of kinematic information for all the individual sources, the best way to isolate cluster stars is to take advantage of the multi-band, high-resolution capabilities of \Euclid and to combine colour information with photometric parameters that are sensitive to the point-like or extended shape of each source.
Figure~\ref{fig:cmd_types} shows the ($\IE-\HE$, $\IE$) CMD of all sources falling in the ERO FoV around NGC~6254. When compared to the CMD shown in Fig.~\ref{fig:cmd}, the much higher complexity of the sampled stellar population becomes evident.
In order to unravel such a complexity, we make use of two of the fit parameters provided by {\tt Sextractor}, called \texttt{SPREAD\_MODEL} and \texttt{CLASS\_STAR}. The former parameter is designed\footnote{see \url{https://sextractor.readthedocs.io/en/latest/Model.html}} such that it equals zero for stars, and deviates from zero for more extended (or more compact) sources. The latter parameter, instead, equals $1$ for sources that are well fit by the adopted PSF model, and is $<1$ for sources with a less satisfactory fit. Figure~\ref{fig:sg} shows the behaviour of these parameters as a function of $\IE$ magnitude. It is immediately evident that stars (yellow symbols with \texttt{CLASS\_STAR}$\sim1$) can be rather well distinguished from other sources by imposing a selection on $\lvert$\texttt{SPREAD\_MODEL}$\rvert<0.01$. What is left out of this selection corresponds either to background galaxies (if \texttt{CLASS\_STAR} is $\ll$1, see the red symbols in Fig.~\ref{fig:cmd_types}) or to saturated stars (if \texttt{CLASS\_STAR}$\sim1$, see blue symbols in Fig.~\ref{fig:cmd_types}).

Among what is selected as point-like sources, the stars located closer to the GC centre, and shown as black symbols in Fig.~\ref{fig:cmd_types}, describe the location of cluster members, whereas the remaining stars, shown as green symbols, represent the numerous Milky Way field stars, composed of a mix of distant halo MS stars and nearby M-dwarfs. 
In this case, the colour information plays the crucial role, and in particular we find that the longest colour baseline ($\IE-\HE$) is the combination of filters that best separates field stars, galaxies, and the GC stellar populations.

In summary, the CMD analysis performed in this section on NGC~6254 provides the ingredients to select likely member GC stars in the best way possible, in the absence of kinematical information. The use of the same \texttt{SPREAD\_MODEL} and \texttt{CLASS\_STAR} parameters, together with a colour--magnitude selection has been effective in the case of NGC~6397 as well, and is adopted hereafter, unless specifically indicated.


\section{Results}\label{sec:results}

\subsection{Surface-brightness profiles}\label{sec:sb}

In order to build a GC surface-brightness profile that is sensitive even to very low stellar surface density, we select a sample of likely cluster member stars by using the photometric parameters described before and by targeting completeness over purity. In this sense, we applied only conservative cuts of \texttt{CLASS\_STAR}$_{\IE}>0.03$, |\texttt{SPREAD\_MODEL}|$_{\IE}< 0.01$, |\texttt{SPREAD\_MODEL}|$_{\HE}< 0.01$, and \texttt{FLUX\_RADIUS}$_{\IE}<1.5$,\footnote{\texttt{FLUX\_RADIUS} is a {\tt Sextractor} parameter that describes the half-light radius of a source light profile.} to exclude obvious artefacts and extended sources. Then, we select our sample of GC stars along a region in the reddening-corrected CMD centred around the theoretical isochrone shown in Fig.~\ref{fig:cmd}, down to $\IE=24$, and in a band $\sim0.3$ mag wide in $\IE-\HE$ colour. As described by Fig.~\ref{fig:cmd_types}, this selection should include the vast majority of GC members.

\begin{figure}[htbp!]
\centering
\includegraphics[width=\columnwidth]{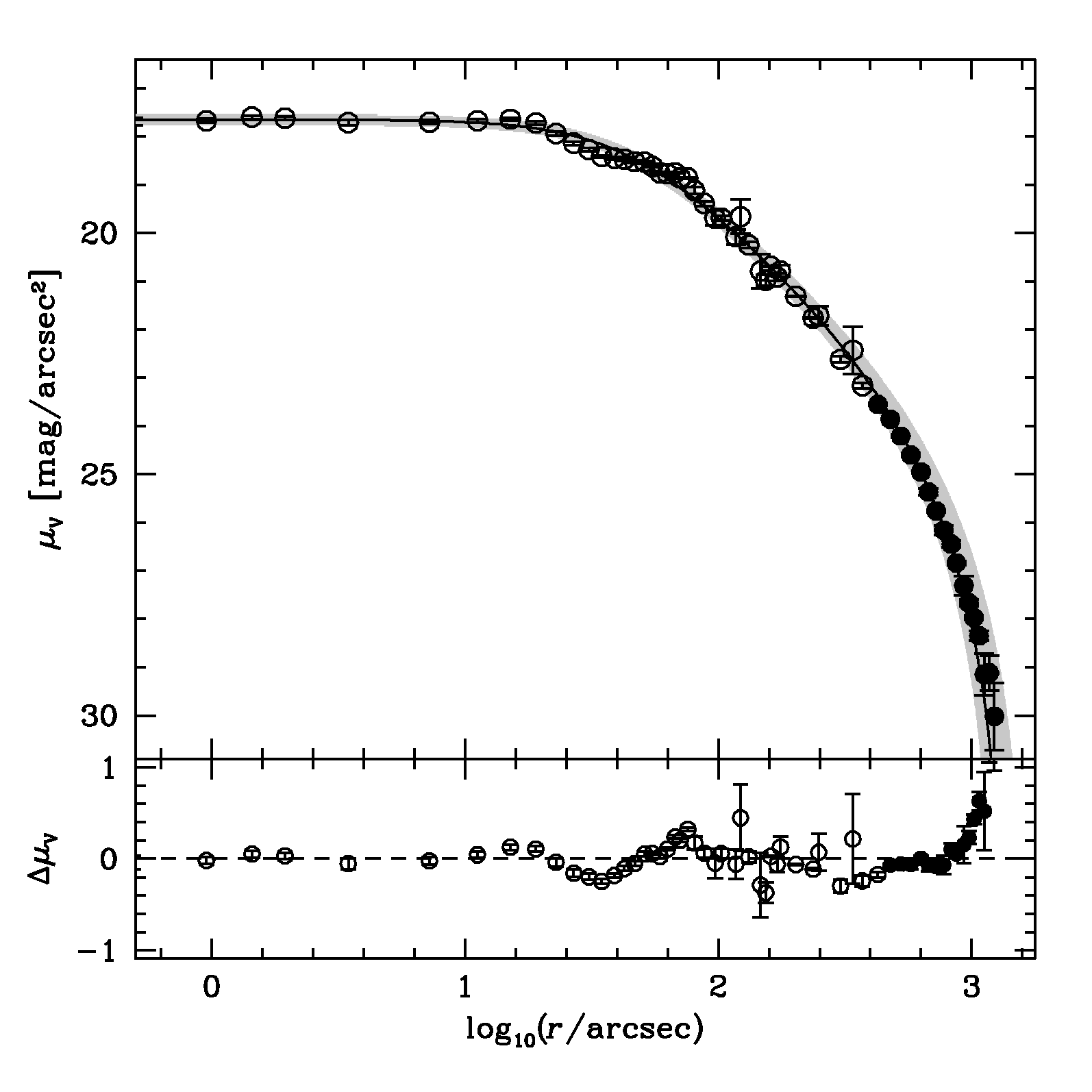}
\caption{Surface brightness profile for NGC~6254. \Euclid data are shown as black, filled symbols, while literature values are represented by open symbols. The black, solid line marks the best-fit King model, and the grey shaded area indicates the associated uncertainty. Finally, the bottom panel shows the residuals of the fit.}
\label{fig:profm10}
\end{figure}

\begin{figure}[htbp!]
\centering
\includegraphics[width=\columnwidth]{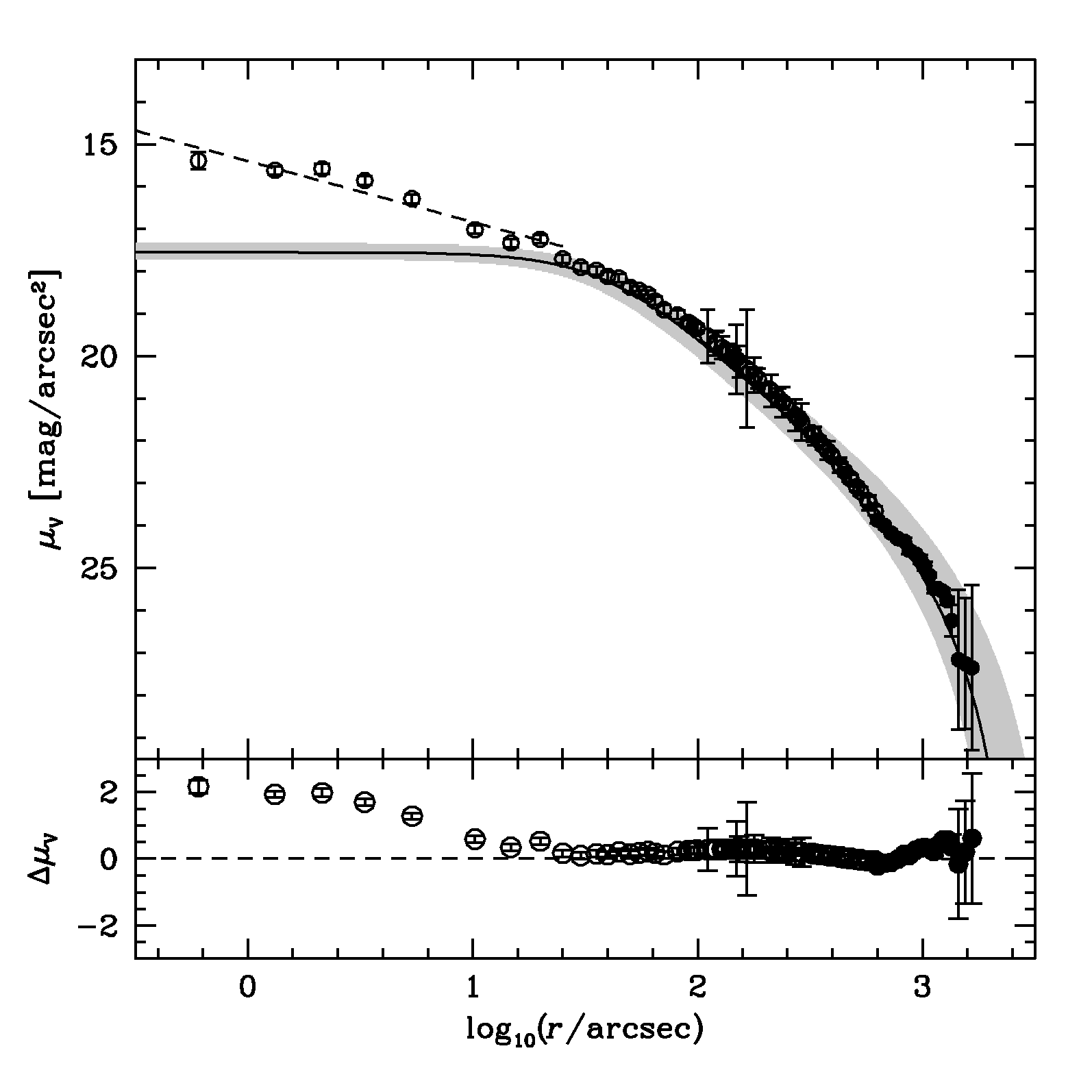}
\caption{Same as in Fig.~\ref{fig:profm10} but for NGC~6397.}
\label{fig:prof6397}
\end{figure}

By using these samples of selected stars, we obtain the surface-density profiles from resolved star counts for both clusters. For NGC~6254, we sub-divided the total FoV in $25$ concentric annuli centred on the cluster's centre of gravity from \citet{dalessandro13}, reaching a maximum cluster-centric distance $r = 1400\arcsec$, while we split the area sampled for NGC~6397 into $28$ annuli centred on the centre reported by \cite{goldsbury10}, thus reaching a maximum radial coverage of $r=1800\arcsec$.  Each annulus was then split in an adequate number of angular sub-sectors (two or four depending on the number of stars).
In each sub-sector, the density has been estimated as the ratio between the selected number of stars counted and the covered area. The density assigned to a given annulus is the average of the densities of each sub-sector of that annulus. The error assigned to each density measure is defined as the dispersion around the mean of the individual sub-sector densities.
We estimated the background density by using the most external data points, which attained almost constant density values. For NGC~6254 we used the six most external density values located at $r>1100\arcsec$, while for NGC~6397 we could only use two data-points at $r>1600\arcsec$. It is worthwhile noticing that, especially in the latter case, a few GC members are still present at these extreme radii, hence the background density might be slightly overestimated.
We then subtracted the mean background values from all the other annuli. Since the \Euclid data are strongly incomplete in the inner cluster regions, we combined our resolved star count profiles with the integrated light surface brightness profiles used by \citet{baumgardt18}.
To this aim, we first reported our density profiles to the same scale and photometric reference systems of the literature profiles, by using values in common in the radial range $250\arcsec<r<400\arcsec$ for NGC~6254 and $400\arcsec<r<650\arcsec$ for NGC~6397. The resulting combined surface density profiles are shown in Fig.~\ref{fig:profm10} and Fig.~\ref{fig:prof6397} for NGC~6254 and NGC~6397, respectively.

The first key result coming from this analysis is that, while sampling almost the same radial range as previous studies, we reach surface brightness values $\mu_V\sim 30.0$ mag~arcsec$^{-2}$ for NGC~6254 and $\mu_V\sim 28$ mag~arcsec$^{-2}$ for NGC~6397, which is more than one magnitude fainter than previously obtained (e.g., \citealt{baumgardt18}). This performance already demonstrates the advance that \Euclid will guarantee on this kind of study in the future. 

We performed a fit of the azimuthally averaged density profiles by using single-mass King models \citep{king62} for both clusters, and show the results of the fit in Fig.~\ref{fig:profm10} and Fig.~\ref{fig:prof6397}. The general approach follows the one described in \cite{dalessandro2013b}. For NGC~6254 the best fit is obtained with a core radius $r_{\rm c}=44.2^{+1.7}_{-1.3}$ arcsec and a concentration parameter $c=1.47^{+0.07}_{-0.05}$, which yields a value for the tidal radius $r_{\rm t}=1375.7^{+216.2}_{-247.3}$ arcsec. These results are in good agreement within the errors with previous estimates in the literature (e.g., \citealt{harris96,dalessandro13}). We note that $r_{\rm t}$ is $\sim20\%$ larger (but still compatible within the errors) than in previous derivations. 
Such a large value can be naturally explained by our use of faint, low-mass MS stars, which tend to preferentially populate the cluster outer regions, while results in the literature are driven by the distribution of heavier red giant branch stars, which are generally more concentrated towards the cluster centre.

For NGC~6397, we excluded from the fit the eight innermost points ($r < 12\arcsec$) of the observed profile, since they appear to deviate from a flat core behaviour, as expected for a post-core collapse cluster.
In this way, the best-fit model is obtained for a core radius $r_{\rm c}=42.2^{+2.5}_{-5.3}$ arcsec, concentration $c = 1.77^{+0.16}_{-0.12}$ and $r_{\rm t}=2580.1^{+1110.3}_{-981.3}$ arcsec. The innermost part of the profile is nicely fit with by a power law with a slope $\alpha\sim0.5$ as typically expected for post-core collapse clusters. Because of the double fit performed in this work, results cannot be directly compared with the literature, where fits have only been performed by using a single King model. 

\subsection{Outer regions morphology}

\begin{figure}[htbp!]
\centering
\includegraphics[scale=0.5]{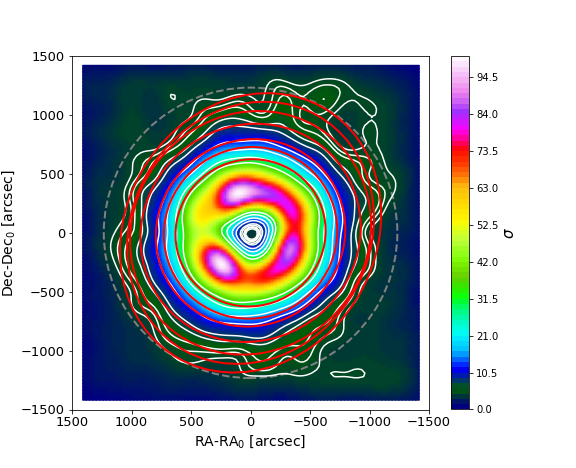}
\caption{Two-dimensional morphology of NGC~6254 likely member stars. The map is colour-coded based on the stellar density, in units of $\sigma$ above the background. Iso-density contours are shown in white. The best-fitting ellipses are over-plotted in red. The tidal radius as determined in this work is marked as a dashed grey circle.} 
\label{fig:2d}
\end{figure}

Given the outstanding performance shown by our ERO data in terms of surface-brightness depth, we use the same sample of member stars described above to look for possible evidence of tidal tails in the two-dimensional morphology of NGC~6254 and NGC~6397. To do so we start from NGC~6254 and compute the stellar-density map shown in Fig.~\ref{fig:2d} with over-plotted density contours at regular density intervals.

As is evident from the figure, the contours are rather circular in the inner regions but, moving outwards, they clearly develop an elongation towards the north-west direction. To provide a quantitative statistical significance of the density contours plotted, we selected members stars in a region more distant than 20 arcmin from the cluster centre and computed the mean stellar density and the dispersion around the mean ($\sigma_{\rho}$) in the cells of a $40\times40$ grid. The lowest-density contour shown in Fig.~\ref{fig:2d} corresponds to a density value of $5.0\,\sigma_{\rho}$.

\begin{figure}[htbp!]
\centering
\includegraphics[width=\columnwidth]{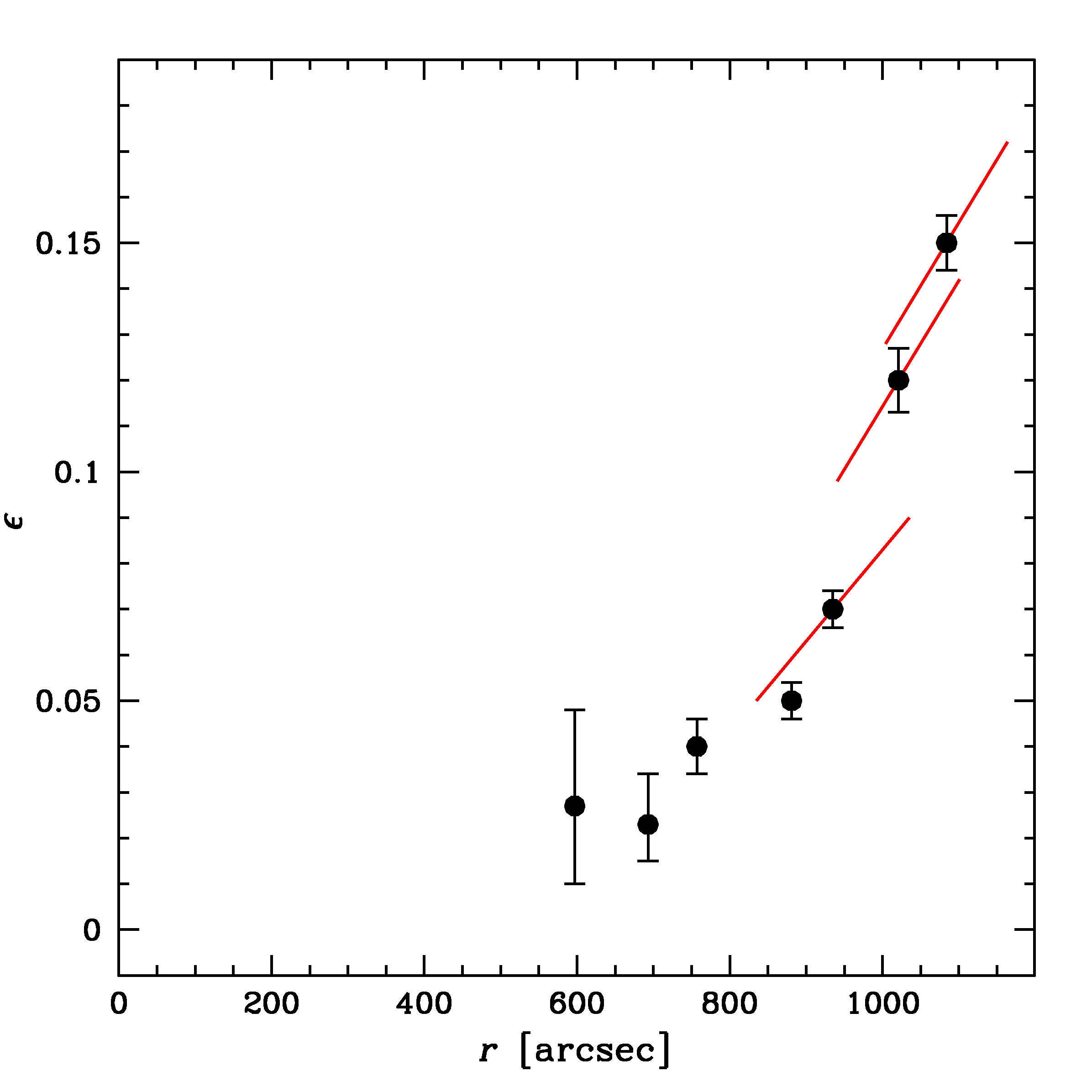}
\caption{Ellipticity of the iso-density contours as a function of distance from the centre of NGC~6254. The red lines indicate the position angle $\Theta$ of the best-fitting ellipse.} 
\label{fig:elli}
\end{figure}

We fit the iso-density contours with ellipses to get a quantitative idea of the morphological elongation in the cluster's outer regions. We fixed the centres of the ellipses on (0,0), and left the angle and the axis ratio as free parameters of the fit. The solutions of the fit are shown in Fig.~\ref{fig:2d} as red ellipses. The associated uncertainties have been estimated by means of a bootstrapping algorithm, repeated 1000 times. For each iso-density contour, we discarded 50\% of the sample of points used to estimate the ellipticity,\footnote{defined as $\epsilon=1-(b/a)$, where $a$ is the semi-major axis and $b$ is the semi-minor axis of the ellipse.} and performed the fit again. The 16th- and 84th percentiles of the resulting distributions of $\epsilon$ are taken as asymmetric error estimates. The ellipticity of the iso-density contours is about zero in the inner regions and then progressively increases towards the GC peripheries, as expected in case of tidally induced distortions, reaching a maximum of $\epsilon=0.15$. This trend is further described in Fig.~\ref{fig:elli}, where the major-axis angles\footnote{defined to increase from west to north, with $\Theta=0$ degrees coinciding with the west direction} $\Theta$ of the ellipses for the contours deviating most from circularity are represented by the tilt shown by the solid red lines, namely $\Theta=60.5$ deg, $\Theta=61.4$ deg and $\Theta=51.6$ deg, for a mean angle of $\ave{\Theta}=57.8$ deg.

\begin{figure}[htbp!]
\centering
\includegraphics[scale=0.5]{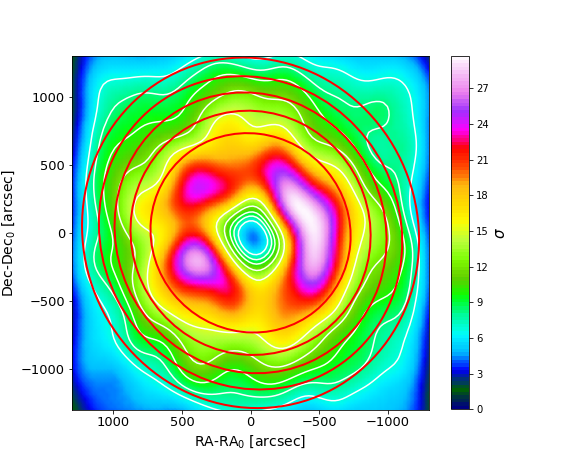}
\caption{Two-dimensional morphology of NGC~6397 likely member stars. The map is colour-coded based on the stellar density, in units of $\sigma$ above the background. Iso-density contours are shown in white. The best-fitting ellipses are over-plotted in red. Note that the tidal radius of NGC~6397 is outside the limits of the figure.} 
\label{fig:2d6397}
\end{figure}

When performing the same analysis on NGC~6397, we obtain the results shown in Fig.~\ref{fig:2d6397}. The first obvious feature is that the contours appear less elongated compared to the case of NGC~6254. Also in this case they appear circular in the inner regions, and then develop a somewhat higher ellipticity towards the cluster periphery, in the north--south direction.

\begin{figure}[htbp!]
\centering
\includegraphics[width=\columnwidth]{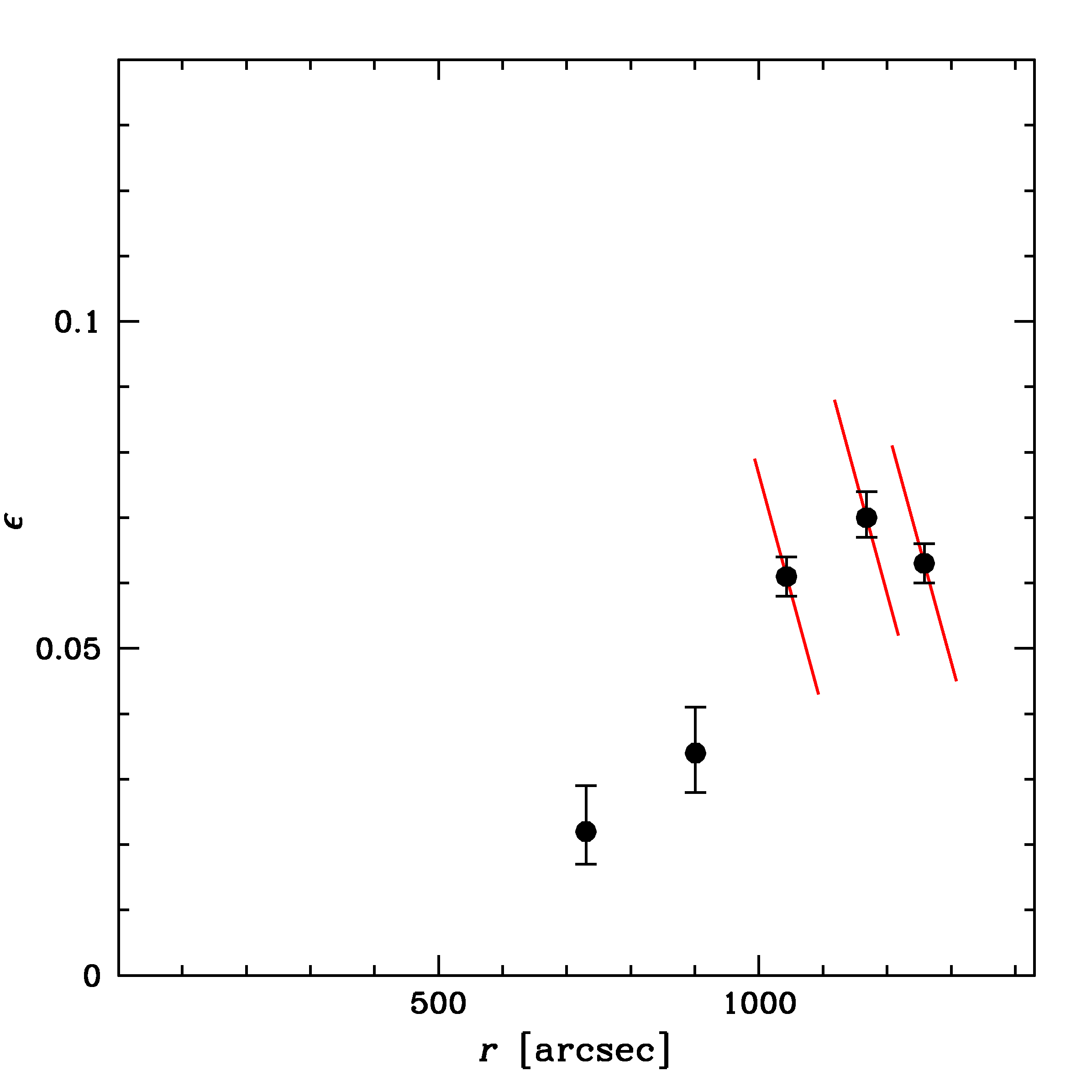}
\caption{Ellipticity of the iso-density contours as a function of distance from the centre of NGC~6397. The red lines indicate the angle of the best-fitting ellipse.} 
\label{fig:elli6397}
\end{figure}

The solution of the ellipsoidal fitting shown in Fig.~\ref{fig:elli6397} confirms the milder morphological elongation. A maximum ellipticity of $\epsilon=0.07$ is found in the two most external contours. 
The fact that we detect a weaker morphological distortion than in NGC~6254 is not surprising. The observed area for NGC~6397 covers a smaller portion of the cluster extent, as NGC~6397 is closer to the Sun and therefore appears more extended on the sky. Our observations can sample only out to about half of the cluster tidal radius of $r_{\rm t}\sim0.8$ deg (see Sect.\,\ref{sec:sb}). This is not likely far enough to detect distortions related to tidally induced effects. It is nonetheless reassuring that the ellipticity we find matches the one quoted by \cite{harris96} and \cite{chen10}. The latter work also quotes a direction of the elongation ($11\pm2$ deg oriented anti-clockwise from the north) that is very similar to ours. In fact, we find that the angle of the best-fit ellipses of the three most elongated density contours of our NGC~6397 map is rather stable, varying between $\Theta=109.5$ deg and $\Theta=114.2$ deg, with a mean value of $\Theta=111.0$ deg, corresponding to $21.0$ deg in the reference system of \cite{chen10}.

\section{Predictions from $N$-body simulations of NGC~6254}\label{sec:nbody}

In order to understand how the morphological features observed in NGC~6254 and NGC~6397 compare with expectations, we perform an $N$-body simulation of each globular cluster in the Milky Way potential. This is done using {\tt gadget-3}, which is an updated version of {\tt gadget-2} \citep{Springel_2005}. For the Milky Way potential, we take the \texttt{MWPotential2014} model from \cite{Bovy_2015}, which consists of a Navarro--Frenk--White dark matter halo \citep{NFW_1997}, a Miyamoto--Nagai disk \citep{Miyamoto-Nagai_1975}, and a broken-power-law bulge. We take the present-day phase-space coordinates for NGC~6254 and NGC~6397 from \cite{Vasiliev_Baumgardt_2021}. Starting from their present-day coordinates, we rewind a tracer particle for 2 Gyr to generate the initial phase-space coordinates for our models. For NGC~6254, we inject a King profile with a concentration parameter $W=7.06$, a tidal radius of $r_{\rm t} = 38.47$ pc, and a mass of $1.5\times10^5 M_\odot$ \citep{de_Boer_etal_2019,McLaughlin_etal_2005}. We model this with $5\times10^5$ equal mass particles and a particle softening of 0.0063\,pc. For NGC~6397, we inject a King profile with $W=8.65$, a tidal radius of $r_{\rm t}=32.16$ pc, and a mass of $1.13\times10^{5} M_\odot$ \citep{de_Boer_etal_2019,Vitral_etal_2022}. We model this with $10^5$ equal mass particles with a particle softening of 0.0063\,pc. Both globular cluster models are then evolved for 2\,Gyr to the present day. We stress that this is an initial $N$-body model of both systems to study their expected tidal debris (i.e. orientation and ellipticity). As such, these models are initialised with the present-day density profiles for each globular cluster and are disrupted for a short amount of time compared to their age ($\sim 13$ Gyr). These models are not designed to faithfully capture important effects such as two-body scattering and mass segregation. Accounting for these effects would require starting with a realistic initial mass function, using a collisional $N$-body code like NBODY6 \citep{Aarseth_1999}, and evolving the system for the age of the cluster, as was done in e.g. \cite{baumgardt18}.

\begin{figure}[htbp!]
\centering
\includegraphics[width=\columnwidth]{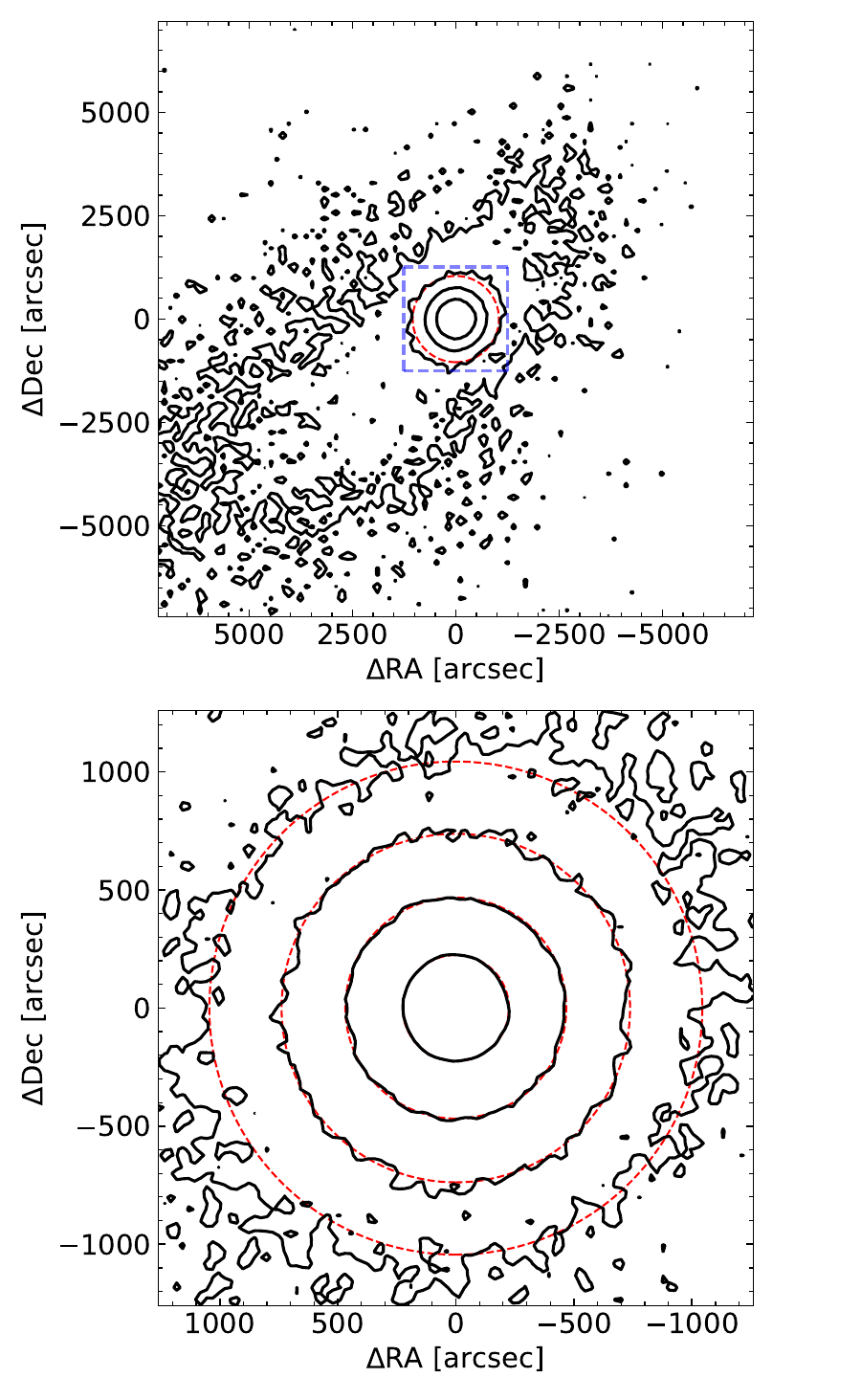}
\caption{Contour maps of an $N$-body simulation of NGC~6254. \textit{Top panel} shows NGC~6254 over a large ($4\times4$ deg$^{2}$) field of view. The dashed-blue square shows the field of view of \Euclid\/. A dashed-red circle with a radius of 1044 arcsec is included to show that the simulated cluster is tidally distorted at this radius. The contours are spaced in multiples of 10. \textit{Lower panel} shows the simulation over the field of view of \Euclid\/ ($0.7\times0.7$ deg$^{2}$). The dashed red circles are chosen to roughly match the contours to show where the cluster is expected to be tidally stretched. The circles are at radii of 225 arcsec, 468 arcsec, 738 arcsec, and 1044 arcsec. The contours are again separated by factors of 10. Note that the largest circle in the bottom panel matches the circle in the top panel.}
\label{fig:nbody_sim}
\end{figure}

\begin{figure}[htbp!]
\centering
\includegraphics[width=\columnwidth]{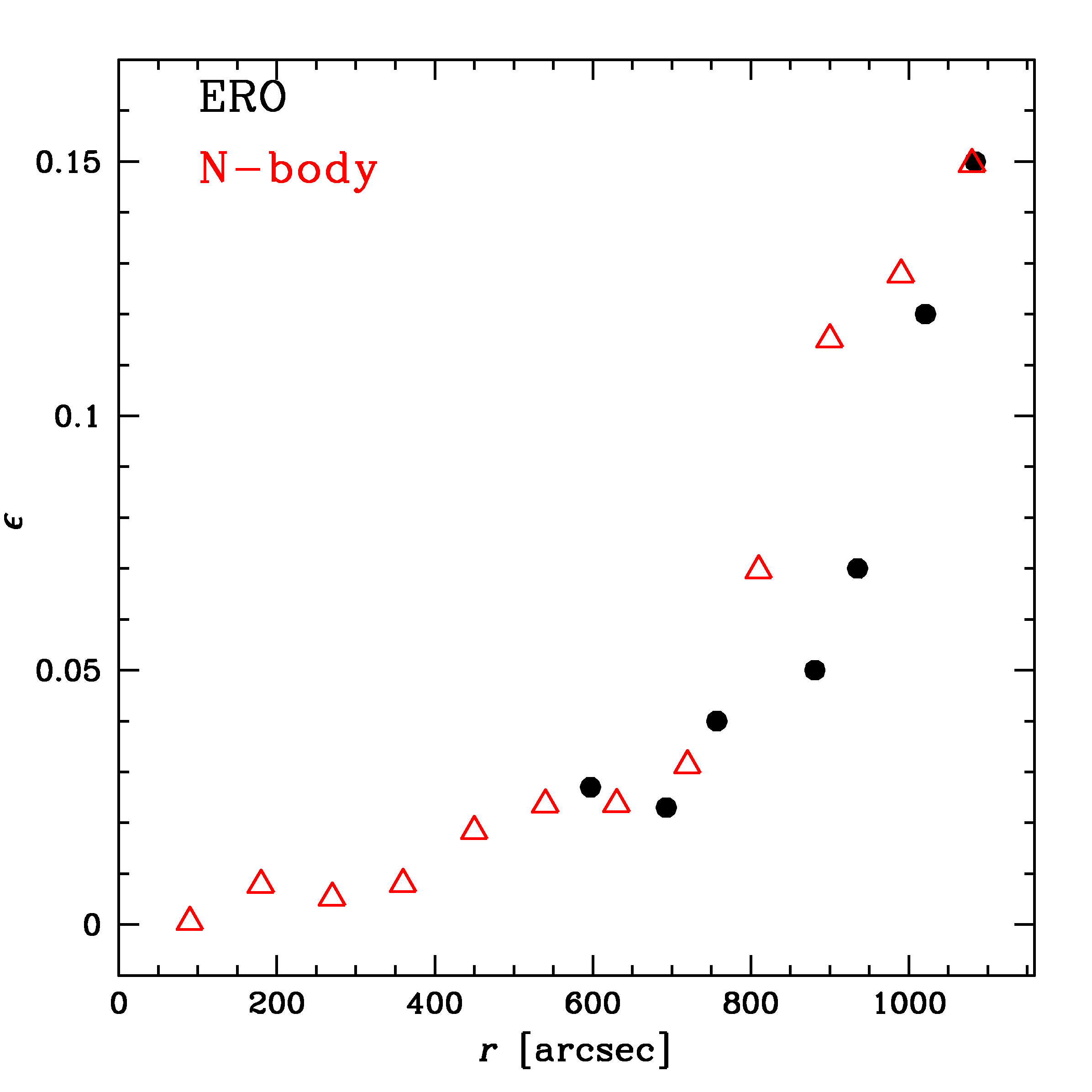}
\caption{Comparison between the radial variation of the ellipticity observed for NGC~6254 (black symbols) and the prediction from $N$-body simulations (red triangles).} 
\label{fig:obsNbody}
\end{figure}

In Fig.~\ref{fig:nbody_sim} we show mock observations of our model of NGC~6254. We note that the simulated cluster is offset from the correct present-day location by $0.08$\,kpc in position, $3.0$\, \kms ~in velocity and $0.67$\,degrees in the plane of the sky from the real NGC~6254. As a result, Fig.~\ref{fig:nbody_sim} shows the coordinates relative to the centre of our simulated cluster. The star counts from the $N$-body simulation of NGC~6254 display a clear elongation in the north-west direction, which coincides with that found in the observations. When comparing this simulation with the iso-density contours shown in Fig.~\ref{fig:2d}, their similarity is remarkable. This is highlighted even more clearly in Fig.~\ref{fig:obsNbody}, where the predicted (red triangles) and the observed (black circles) variation of NGC~6254 ellipticity as a function of the cluster radius are compared directly. In both cases, the cluster morphology remains about circular in the inner regions and then starts deviating significantly at about $13.5$\,arcmin. The ellipticity increases rather sharply in both the observations and the simulation by moving farther out and reaches its maximum of $\epsilon=0.15$ at $r\sim18.5$ arcmin. We note that in the $N$-body simulation, the ellipticity is estimated using the eigenvalues of the reduced moment of inertia tensor formed out of the on-sky positions of the particles in circular annuli. 

The consistency between the predictions from the $N$-body simulations strengthens significantly the interpretation of the morphological
distortion we detect in the outer regions of NGC~6254 as an actual feature. This clearly indicates that NGC~6254 has experienced tidal forces due to its interaction with the Milky Way, and that these forces have distorted the cluster morphology, likely leading to the development of tidal tails on larger scales (outside the GC Jacobi radius).

\begin{figure}[htbp!]
\centering
\includegraphics[width=\columnwidth]{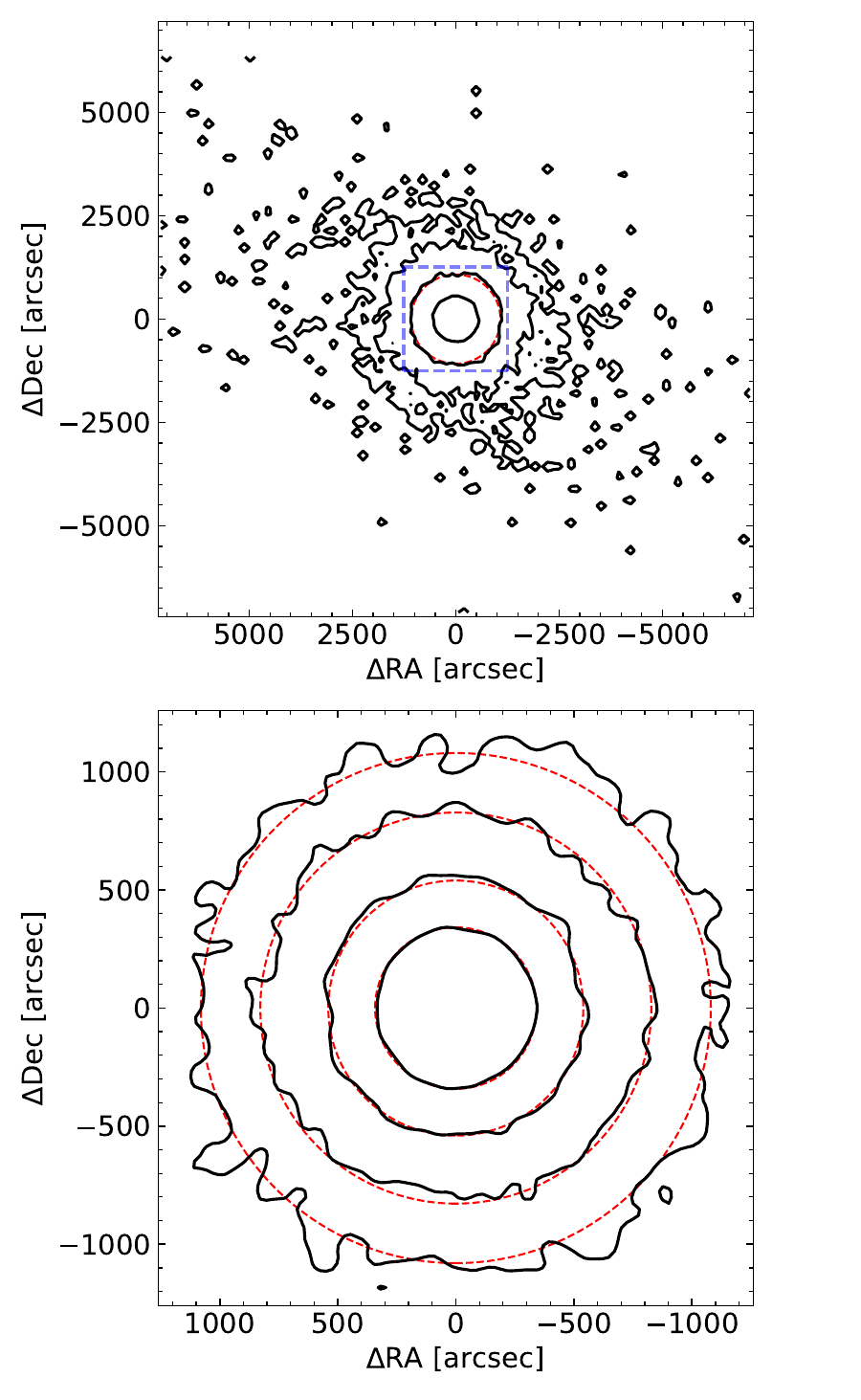}
\caption{Contour maps of an $N$-body simulation of NGC6397. \textit{Top panel} shows NGC~6397 over a large ($4\times4$ deg$^{2}$) field of view. The dashed-blue square shows the field of view of \Euclid. A dashed-red circle with a radius of 1080 arcsec is included to show that the simulated cluster is not significantly tidally distorted at this radius. The contours are spaced in multiples of 10. \textit{Lower panel} shows the simulation over the field of view of \Euclid. The dashed-red circles are chosen to roughly match the contours to show where the cluster is expected to be tidally stretched. The circles are at radii of 342 arcsec, 540 arcsec, 828 arcsec, and 1080 arcsec. The contours are separated by half a dex. Note that the largest circle in the bottom panel matches the circle in the top panel.} 
\label{fig:nbody_sim_6397}
\end{figure}

In Fig.~\ref{fig:nbody_sim_6397} we show mock observations of our model of NGC~6397. As with NGC~6254, our $N$-body simulation does not exactly match the present-day location and differs by $0.19$\,kpc, $12$\,\kms, and $1.1$\,degrees on the sky with observations of NGC~6397. As a result, the on-sky angles shown in Fig.~\ref{fig:nbody_sim_6397} are relative to the centre of the simulated cluster. Our $N$-body simulation of NGC~6397 shows no significant tidal deformation within the \Euclid field of view, but indicates that tidal deformations should be visible on larger scales. This is consistent with the fact that the Jacobi radius of NGC~6397 is larger than the FoV of \Euclid.

\section{Discussion and conclusions}\label{sec:concl}

In this paper, we analyse \Euclid ERO data of two Milky Way GCs, namely NGC~6254 and NGC~6397, to investigate the presence of tidally induced morphological distortions caused by the clusters' interaction with the Milky Way.

As a first result of our analysis, we have shown the first ever \Euclid CMDs of Galactic GCs. The precision in colour, $\lesssim0.04$ mag, seems to indicate the presence of a marginal colour spread in the low-mass regime, that could be associated with GC multiple populations \citep[e.g.,][]{milone19} if confirmed by future more detailed studies. The unprecedented combination of wide-field capability and depth further enables the detection of faint, low-mass stars down to $~0.15$ M$_{\odot}$ and out to the GC tidal radius, at least in the case of NGC~6254. Improved understanding of the data and of the photometric analysis in the future will enable sampling even lower stellar masses.

The surface brightness profiles built from accurately selected samples of cluster members reach more than one magnitude fainter than the surface brightness profiles presented in the literature so far for these two GCs \citep[see e.g.,][]{baumgardt18}, down to a value of $\mu_V\simeq30.0$\,mag~arcsec$^{-2}$. Such a faint limit makes them amongst the deepest ever determined, and the first to be constructed using star counts at the bottom of the MS. The tidal radius estimated from the best-fit King model of NGC~6254 is $\sim20\%$ larger than previous estimates, and such an increase can be explained by the fact that \Euclid enables, for the first time, the use of low-mass MS stars, rather than giants, as morphological tracers.

Most importantly, the findings of this work provide the first robust evidence for tidally induced morphological distortions in the outer regions of NGC~6254, which implies a high chance of finding tidal tails around the cluster on the larger scales, which will be covered by the \Euclid survey.

This evidence comes from the analysis of the two-dimensional morphology of the cluster. In fact, iso-density contours built from a sample of likely member stars reveal an elongated morphology of NGC~6254 outer regions at more than a $5\,\sigma$ significance. Such an elongation is oriented along the north-west direction with a mean angle of about $58$ deg, which coincides remarkably well with the direction of the cluster tidal distortions predicted by $N$-body simulations, and disappears towards the cluster inner regions. It should be noted that the direction of the elongation is somewhat reminiscent of the direction in which the extinction varies through the FoV (see Fig.~\ref{fig:extinction}). However, our selection of likely cluster members has been performed on a differential reddening-corrected CMD, where the limiting magnitude is the one in the $\HE$ band. The colour-excess variation $\delta E(B-V)$ across the field amounts to $0.15$ mag corner-to-corner, which based on the extinction law by \cite{cardelli89} corresponds to a difference in depth of only 0.08 mag in $\HE$. We can thus safely conclude that the detected morphological distortion is not induced by differential extinction-related effects.
Moreover, we presented a remarkable similarity between the ellipticity gradient observed from the data and the one predicted by the $N$-body simulations. Both start from an almost circular morphology in the inner $\sim10$ arcmin, and then rapidly increase to the maximum value of $\epsilon=0.15$ at a distance of $\sim18.5$ arcmin. This further supports the notion that the morphological elongations are indeed due to tidal features.

Finally, the two-dimensional morphology of NGC~6397 could only be sampled out to half of the GC tidal radius, and reveals a milder elongation ($\epsilon_{\rm max}=0.07$) along about the north-south direction, in agreement with previous studies. Further investigation on larger scales, sampling regions further out than the cluster tidal radius ($r_{\rm t}\simeq0.8$ deg as found in Sect.\,\ref{sec:sb}), is required to more robustly interpret this feature in terms of tidally induced morphological distortion.

\begin{figure*}[btp!]
\centering
\includegraphics[width=\textwidth]{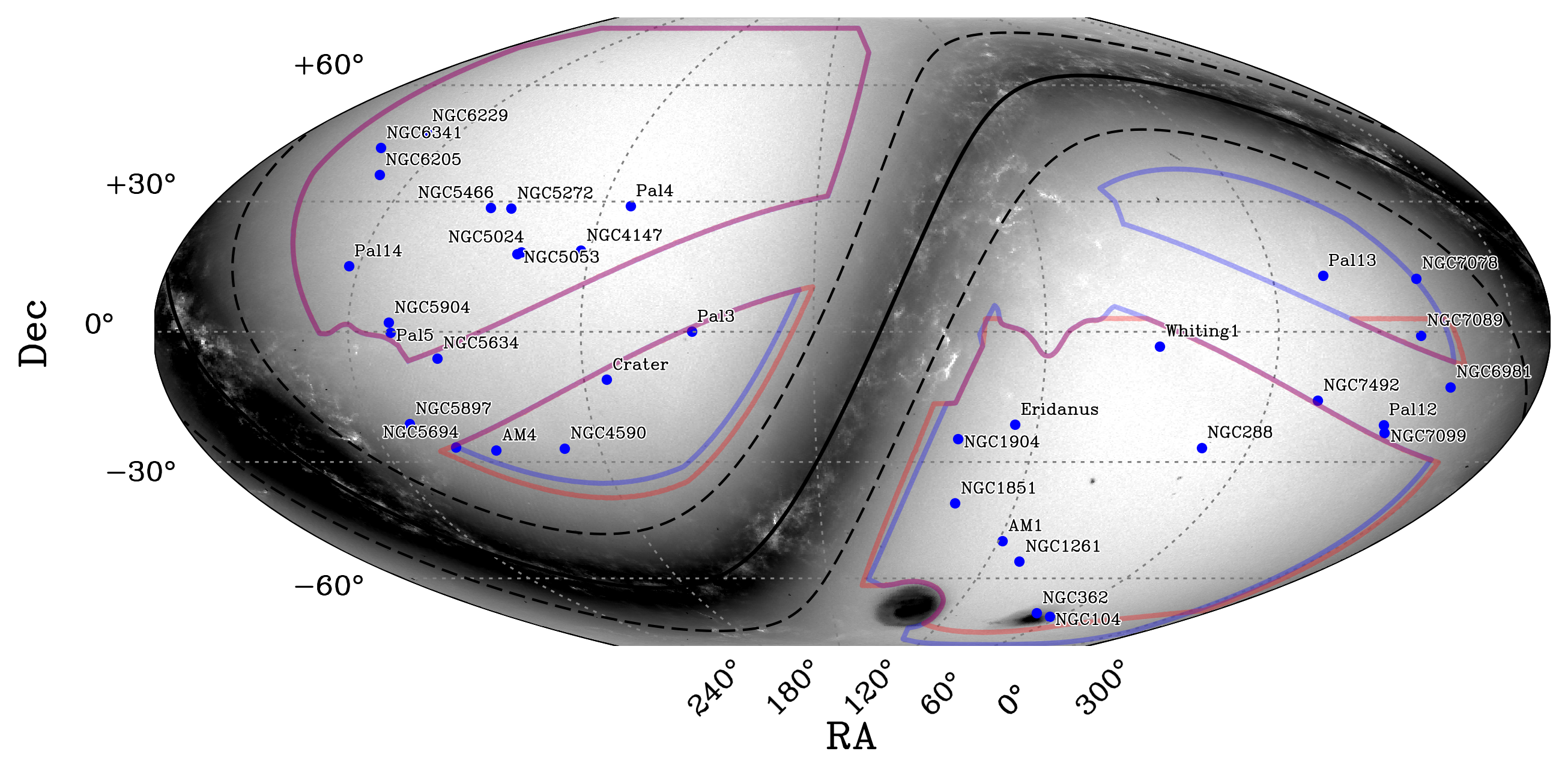}
\caption{Planned Region of Interest (blue polygons) and footprint (red polygons) of the full 6 year \Euclid wide survey shown in an McBryde-Thomas Flat Polar Quartic projection within blue boxes. The background represents {\it Gaia} DR3 number counts in logarithmic scale. The position and the names of the GCs covered by the survey or located in its proximity are highlighted in blue. } 
\label{fig:survey}
\end{figure*}

To conclude, the findings of this \Euclid ERO project demonstrate the potential that the \Euclid survey has for advancing the study of star clusters in our Galaxy. To determine which GCs fall within the nominal \Euclid\/ survey coverage, we matched the 6 year \Euclid Wide Survey Region of Interest \citep{Scaramella-EP1} and planned footprint\footnote{We remark that the footprint currently shown in Fig.~\ref{fig:survey} could be subject to minor changes}, shown in Fig.~\ref{fig:survey} as blue and red polygons, respectively, with a table of the known MW GCs from \cite{baumgardt18}. In total, we find that more than $\sim20$ MW GCs will fall in the planned \Euclid survey footprint. For all of these, the search for tidal tails, as well as the investigation of many other GC-related science cases, can be carried out successfully with \Euclid. 

\begin{acknowledgements}
  We are grateful to the anonymous referee for the invaluable feedback that improved the quality of the paper. DM thanks the Fundacti\'on Jes\'us Serra visiting programme and the Instituto de Astrof\'isica de Canarias for hospitality. DM acknowledges financial support from the European Union – NextGenerationEU RRF M4C2 1.1  n: 2022HY2NSX. "CHRONOS: adjusting the clock(s) to unveil the CHRONO-chemo-dynamical Structure of the Galaxy” (PI: S. Cassisi).  
  ED aknowledges financial support from the Fulbright Visiting Scholar program 2023. ED is also grateful for the warm hospitality of the Indiana University where part of this work was performed.  IM acknowledges funding from UKRI/STFC through grants ST/T000414/1 and ST/X001229/1, and has received funding from the European Union’s Horizon 2020 research and innovation programme under grant agreement No 101004214. AL acknowledges funding from Agence Nationale de la Recherche (France) under grant ANR-19-CE31-0022
\AckERO \AckEC
\end{acknowledgements}

%
%
\bibliography{Euclid}

%
%

%

\end{document}